\documentclass{emulateapj}
\newcommand{\p}{\partial}
\newcommand{\e}{{\rm e}}
\newcommand{\ev}{{\rm ev}}
\newcommand{\dd}{{\rm d}}
\newcommand{\ie}{i.e. }
\newcommand{\eg}{e.g. }
\newcommand{\etal}{et al. }

\newcommand{\M}{{\cal M}}

\newcommand{\sh}{{\rm sh}}
\newcommand{\Mc}{{{\cal M}^2}}
\newcommand{\hh}{\delta h}
\newcommand{\z}{z}

\newcommand{\inn}{{\rm in}}
\newcommand{\out}{{\rm out}}
\newcommand{\Deltaz}{H}

\begin{document}

\title{A simple toy model of the advective-acoustic instability I. Perturbative approach}
\author{T. Foglizzo$^{1,2}$}
\affil {$^{1}$ CEA, Irfu, SAp, Centre de Saclay, F-91191 Gif-sur-Yvette, France.\\
$^{2}$ UMR AIM, CEA-CNRS-Univ. Paris VII, Centre de Saclay, F-91191 Gif-sur-Yvette, France.
}
\email{foglizzo@cea.fr}

\begin{abstract}
Some general properties of the advective-acoustic instability are described and understood using a toy model which is simple enough to allow for analytical estimates of the eigenfrequencies. 
The essential ingredients of this model, in the unperturbed regime, are a stationary shock and a subsonic region of deceleration. For the sake of analytical simplicity, the 2D unperturbed flow is parallel and the deceleration is produced adiabatically by an external potential. 
The instability mechanism is determined unambiguously as the consequence of a cycle between advected and acoustic perturbations. The purely acoustic cycle, considered alone, is proven to be stable in this flow. Its contribution to the instability can be either constructive or destructive. A frequency cut-off is associated to the advection time through the region of deceleration. 
This cut-off frequency explains why the instability favours eigenmodes with a low frequency and a large horizontal wavelength. The relation between the instability occurring in this highly simplified toy model and the properties of SASI observed in the numerical simulations of stellar core-collapse is discussed. This simple set up is proposed as a benchmark test to evaluate the accuracy, in the linear regime, of numerical simulations involving this instability. We illustrate such benchmark simulations in a companion paper.
\end{abstract}


\section{Introduction}

In the scenario proposed by Bethe \& Wilson (1985) for core collapse supernovae, the success of the explosion depends on the efficiency of energy deposition by neutrinos below the stalled accretion shock, during the first second after core bounce. Numerical simulations revealed that this mechanism is inefficient in spherical symmetry  (Liebend\" orfer et al. 2001). Hydrodynamical instabilities may play an important role by breaking the spherical symmetry and helping the revival of the stalled shock. Indeed, multidimensional simulations allowing for transverse motions, induced by neutrino-driven convection in the gain region, approached the explosion threshold (Burrows et al. 1995, Janka \& M\"uller 1996), although this effect did not seem sufficient (Buras et al. 2003).
The discovery of another hydrodynamical instability, named the Standing Accretion Shock Instability (SASI) by Blondin et al. (2003), opened new perspectives which seem very promising for the revival of the shock. The recent simulations of Marek \& Janka (2007) showed a successful explosion of a $15 M_{\sun}$ progenitor where neutrino energy deposition is efficient enough owing to the effect of SASI, which lengthens the time spent by the postshock gas in the gain region (Murphy \& Burrows 2008).
SASI is also the starting point of the acoustic mechanism found by Burrows et al. (2006, 2007), where  the excitation of g-mode oscillations inside the proto-neutron star triggers the emission of acoustic waves (see also Weinberg \& Quataert 2008). 
Besides the question of the explosion mechanism, the development of SASI seems to have important consequences on the birth conditions of the neutron star, its kick (Scheck et al. 2004, 2006) and also its spin (Blondin \& Mezzacappa 2007, Yamasaki \& Foglizzo 2008). \\ 
In view of the spectacular possible consequences of SASI, a fundamental understanding of its mechanism is desired but still a source of debate. It is definitely distinct from convection since it can take place even without any source of heating (Blondin et al. 2003). The ``advective-acoustic cycle'' (Foglizzo \& Tagger 2000, Foglizzo 2001, Foglizzo 2002) was recognized as the driving mechanism by several authors (Blondin et al. 2003, Burrows et al. 2006, Ohnishi et al. 2006, Foglizzo et al. 2007, Scheck et al. 2008, Yamasaki \& Foglizzo 2008), but some alternate interpretations were proposed (Blondin \& Mezzacappa 2006, Blondin \& Shaw 2007). It should be noted that the analytical arguments of Laming (2007) supporting the existence of a purely acoustic mechanism contained some errors, and the corrected calculation favours the advective-acoustic mechanism (Laming 2008, in press). Part of the difficulty in recognizing the advective-acoustic mechanism in numerical simulations, even in the simplified set up proposed by Blondin et al. (2003), comes from the lack of simple reference models where its properties would be fully understood. 
The present work aims at providing such a reference, as simple as possible. This reference serves two purposes. It can help us build our physical intuition about the advective-acoustic coupling responsible for an unstable advective-acoustic cycle. It can also be used as a benchmark test to evaluate the accuracy of multidimensional numerical simulations in the linear phase of the instability (Sato et al. 2008, hereafter paper~II).\\
Although some of the equations used in the present work have already been described in past publications dealing with more complex flows (e.g. Yamasaki \& Foglizzo 2008 and references therein), we choose to explain them again in the simpler framework considered here. 
Analytical calculations are made possible by three main simplifications: (i) the flow is 2D planar, (ii) the region of deceleration is localized spatially at a distance from the shock, and (iii) the deceleration is produced adiabatically. The simplicity of the present set up allows for an exact calculation, even at low frequency. The physical insight in the mechanism at work can be used as a reference, in order to analyze more realistic situations.\\
The paper is organized as follows: in Sect.~2, the toy model consisting of a stationary shock in an external step-like potential is described, and its eigenspectrum is computed as an illustrative example. Sect.~3 explains the method used to prove unambiguously the mechanism at work, beyond the determination of the eigenfrequencies. This method requires the determination of the efficiency of the advective-acoustic coupling at the shock, in Sect.~4, and in the region of acoustic feedback in 
Sect.~5. These local wave coupling efficiencies are combined into a global cycle whose growth rate is computed in Sect.~6 and compared to the direct eigenspectrum of Sect.~2. The importance of the size of the deceleration region is emphasized in both Sect.~5 and 6. The relevance of these results to the problem of core-collapse is discussed in Sect. 7. Results are summarized in Sect.~8. Lengthy analytical derivations are described in Appendices for the sake of clarity.

\section{Description of a toy model\label{Sect_toy}}

\begin{figure}
\begin{center}
\includegraphics[width=\columnwidth]{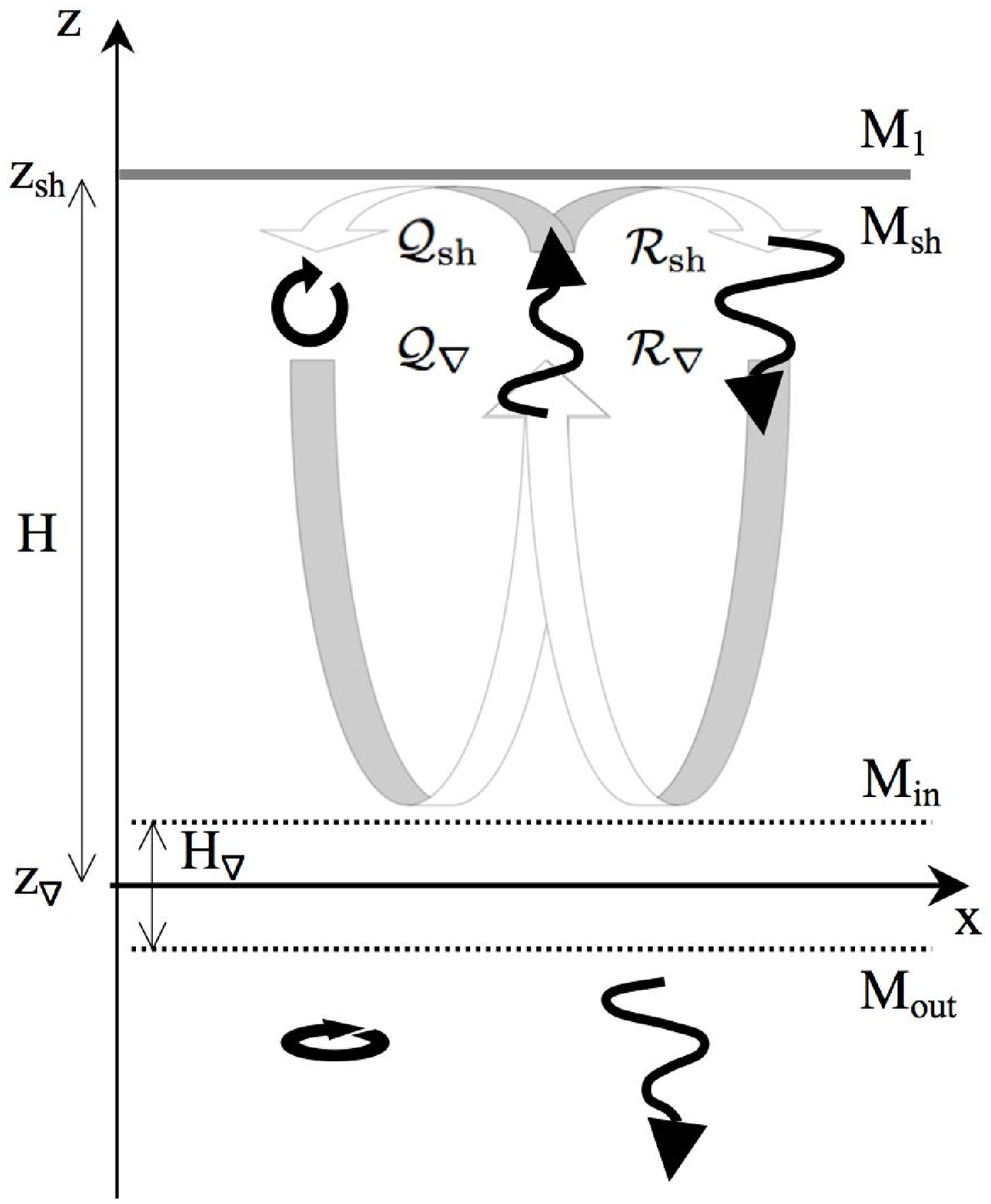}
\caption[]{Schematic view of the toy model. Entropy/vorticity perturbations (circular arrows) are advected downward with the flow, and coupled to acoustic ones (wavy arrows) in the inhomogeneous region near $z_{\nabla}$. The linear coupling between these perturbations is described by coupling coefficients ${\cal Q}_{\rm sh}$, ${\cal Q}_{\nabla}$, ${\cal R}_{\rm sh}$, ${\cal R}_{\nabla}$, defined in Sect.~ 3}
\label{figtoy}
\end{center}
\end{figure}

\subsection{Stationary flow}

The essential ingredient of the advective-acoustic instability is the interplay of advected and acoustic perturbations in a subsonic cavity (\eg the sketch
drawn in Fig. ~1 of Scheck et al. 2008). For the sake of simplicity, we build a toy model where the advective-acoustic cycle takes place without the complications associated with the spherical geometry, and without the difficulties relative to non adiabatic processes. The discussion of these difficulties is postponed to Sect.~7.
We choose the 2D parallel flow of an ideal gas in a Cartesian geometry, passing through a shock and adiabatically decelerated by an external potential. The density is uniform in the $x$ direction and the gas flows along the $z$ direction with a negative velocity. The initial position of the stationary shock is noted $z_{\sh}$ and the region of deceleration centered on $z_\nabla$ (Fig. \ref{figtoy}). The distance from the shock to the center of the deceleration region is noted $\Deltaz\equiv z_{\rm sh}-z_\nabla$.

In the adiabatic approximation, the equations describing the stationary flow of ideal gas, with a density $\rho$ and a velocity $v$ in the external potential $\Phi(z)$, are the conservation of mass flux and the Euler equation:
\begin{eqnarray}
{\p \over\p \z}(\rho v) &=&0,\label{eqcont}\\
{\p\over\p \z}\left({v^2\over 2}+{c^2\over\gamma-1}+\Phi\right)&=&
0,\label{eqbern}
\end{eqnarray}
where $\gamma$ is the adiabatic index, and the sound speed $c$ is related to the pressure $p$ by $c^2\equiv \gamma p/\rho$. The only gradients in this stationary flow are produced by the external potential $\Phi(z)$, localized in the vicinity of the coordinate $z_\nabla$ over a region of size $\Deltaz_\nabla$:
\begin{equation}
\Phi(z)\equiv{\Delta \Phi\over2}
\left\lbrack
\tanh \left({z-z_{\nabla}\over\Deltaz_{\nabla}/2}\right)+1
\right\rbrack.
\end{equation}
At a distance exceeding $\sim 3\Deltaz_\nabla$ from $z_\nabla$, the flow is approximately uniform ($1-\tanh(6) \sim 1.2\times10^{-5}$).
Quantities in the uniform subsonic regions upstream and downstream of the potential jump are denoted by the subscripts ``in" and ``out" respectively, while the subscripts ``1" and ``sh" denote quantities immediately ahead and after the stationary shock respectively. 
The entropy $S$ is defined by
$S\equiv \left\lbrack\log((p/p_{\rm sh})/(\rho/\rho_{\rm sh})^\gamma)\right\rbrack/(\gamma-1)$, and the Mach number is defined as positive: $\M\equiv-v/c$. The only dimensionless parameters of the stationary flow are the adiabatic index $\gamma$, the incident Mach number $\M_1$, the relative size of the coupling region $\Deltaz_{\nabla}/\Deltaz$ and the adiabatic heating parameter $c^2_{\rm out}/c^2_{\rm in}$ (or the deceleration parameter $v_{\rm out}/v_{\rm in}$). Note that $\gamma=4/3$ and $\M_1=5$ throughout the paper.
The potential jump $\Delta\Phi$, the sound speed, velocity, and Mach number jumps are related to the parameters of the flow, through conservation equations, as follows:
\begin{eqnarray}
\M_{\rm in}&=&\left\lbrack{2+(\gamma-1)\M_1^2\over 2\gamma\M_1^2-\gamma+1}\right\rbrack^{1\over2},\\
{\M_{\rm out}\over\M_{\rm in}}&=&\left({v_{\rm out}\over v_{\rm in}}\right)^{\gamma+1\over2},\label{Moutin}\\
{c_{\rm out}\over c_{\rm in}}&=&\left({v_{\rm in}\over v_{\rm out}}\right)^{{\gamma-1}\over2},\\
\Delta\Phi&=&\left({\M_{\rm out}^2\over2}+{1\over\gamma-1}\right)c_{\rm out}^2-
\left({\M_{\rm in}^2\over2}+{1\over\gamma-1}\right)c_{\rm in}^2.
\end{eqnarray}

\subsection{Linear perturbations}

The 1D stationary flow is perturbed in the plane ($x,z$), with periodic boundary conditions in the direction $x$. For a given horizontal size $L_x$ of the computation domain, the wavenumber $k_x$ is restricted to discrete values associated to the number $n_x$ of horizontal wavelengths, $k_x\equiv{2\pi n_x/ L_x}$. The incoming supersonic flow is not perturbed. For the sake of physical simplicity, the lower boundary condition is defined by the free leakage of perturbations, i.e. the absence of an acoustic flux propagating upward from $z<z_\nabla-3\Deltaz_\nabla$. Instead of repeating the tedious derivation of the differential system governing the evolution of linear perturbations, we deduce it from previous studies such as Foglizzo et al. (2006, hereafter FSJ06) or Yamasaki \& Foglizzo (2008), by neglecting non adiabatic effects or rotation. The three functions chosen to describe the evolution of perturbations are noted $\delta f$, $\delta h$ and $\delta S$, where $\delta S$ is the entropy perturbation, $\delta f$ is the the perturbation of the energy density associated to the Bernoulli invariant, and $\delta h$ is associated to the vertical mass flux:
\begin{eqnarray}
\delta f&\equiv&v_z\delta v_z+{2\over\gamma-1}c\delta c\, ,\\
\delta h&\equiv&{\delta v_z\over v_z}+{\delta\rho\over\rho},
\end{eqnarray}
The perturbations of velocity, sound speed, density, pressure, vorticity are directly related to $\delta f,\delta h,\delta S$ through a set of equations recalled in Appendix ~A (Eqs.~(\ref{dvv}-\ref{vorticityy})). \\
The differential system is a simplified version of Eqs. (22-24) of FSJ06:
\begin{eqnarray}
{\p \delta f\over\p z}&=&{i\omega v\over 1-\Mc}\left\lbrack
 \delta h - {\delta f\over c^2}+\left({1\over\M^2}+\gamma-1\right){\delta S\over\gamma}
 \right\rbrack,\label{dfdr1}\\
{\p{ \delta h}\over\p z}&=&{i\omega\over v(1-\Mc)}\left\lbrack
\left(1-{\omega_{\ev}^2\over\omega^2}\right){\delta f \over c^{2}}
 -\Mc { \delta h}- \delta S\right\rbrack
 \label{dhp},\\
{\p \delta S\over\p z}&=&{i\omega\over v}\delta S, \label{dsp}
\end{eqnarray}
where the frequency $\omega_{\ev}$, related to the evanescent or propagating nature of acoustic waves in the vertical direction, is defined by:
\begin{eqnarray}
\omega_{\ev}&\equiv&k_x c(1-\M^2)^{1\over2}.\label{omegaz}
\end{eqnarray}
The boundary conditions at the shock surface are deduced from Eqs.~(28-30) of FSJ06:
\begin{eqnarray}
{\delta f_{\rm sh}\over\omega}&=&iv_1{\Delta \zeta} \left(1-{v_{\rm sh}\over v_{1}}\right) ,\label{bc1}\\
\delta h_{\rm sh}&=&-i{\omega\over v_{\rm sh}}{\Delta \zeta} \left(1-{v_{\rm sh}\over v_{1}}\right) ,\\
{\delta S_{\rm sh}\over\gamma}&=&
i{\omega v_1\over c^{2}}{ \Delta \zeta }\left(1-{v_{\rm sh}\over v_1}\right)^{2}, \label{Ssh}
\end{eqnarray}
where the velocity of the shock is related to its displacement through $\Delta v\equiv -i\omega\Delta\zeta$. 
The expression of the leaking lower boundary condition is deduced from Eq.~(34) in FSJ06, which is based on the decomposition of a perturbation onto acoustic and advected contributions (Eqs.~(\ref{deffS})-(\ref{defhpm}) in Appendix~A):
\begin{eqnarray}
{\mu_{\rm out}\over \M_{\rm out}}{\delta f_{\rm out}\over c_{\rm out}^2} 
-\left(\gamma+{\mu_{\rm out}\over\M_{\rm out}}\;\;{1-\M_{\rm out}^2\over1+\mu_{\rm out}\M_{\rm out}}\right)
{\delta S_{\rm out}\over\gamma}\nonumber\\
- \hh_{\rm out}=0.\label{bcleaking}
\end{eqnarray}
The parameter $\mu$ is defined by a square root of a complex number. The choice of its sign is set by the requirement that the amplitude of acoustic waves $\propto\exp(ik_z^\pm z)$ is bounded in the direction of their propagation (Eq.~(C19) of FSJ06):
\begin{eqnarray}
\mu&\equiv&\left(1-{\omega_{\ev}^2\over\omega^2}\right)^{1\over2},\\
k_z^\pm&=&{\omega\over c}\;\;{\M\mp\mu\over1-\M^2}.\label{ky}
\end{eqnarray}

\subsection{Illustrative examples as possible benchmark tests}

\begin{figure}
\begin{center}
\includegraphics[width=\columnwidth]{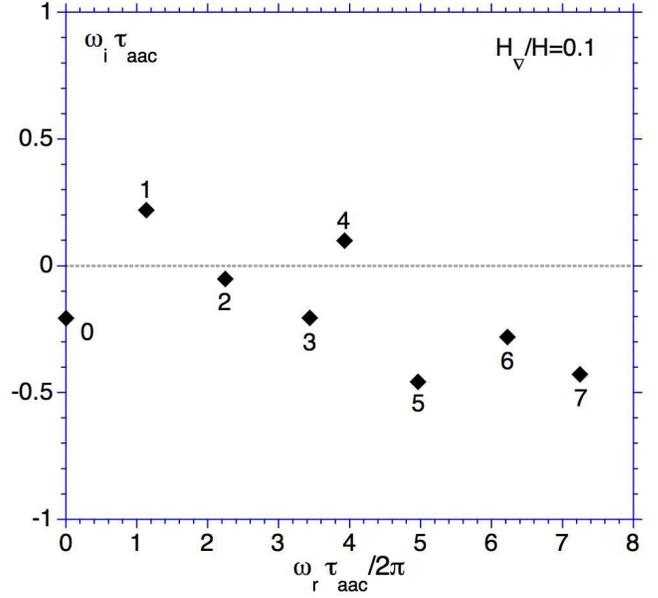}
\caption[]{Eigenfrequencies of the most unstable modes for $0\le n_x\le 7$ in a flow where $\Deltaz_{\nabla}/\Deltaz=0.1$. The value $n_x$ of the corresponding wavenumber is indicated next to each eigenfrequency. Positive values of $\omega_i$ correspond to growth.}
\label{figeigen1}
\end{center}
\end{figure}

\begin{figure}
\begin{center}
\includegraphics[width=\columnwidth]{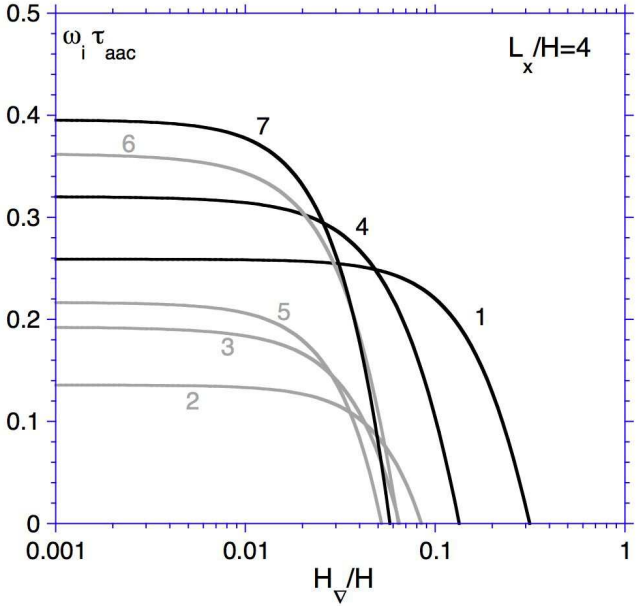}
\caption[]{Dependence of the growth rate $\omega_i$ on the size $\Deltaz_{\nabla}$ of the coupling region, when the amplitude $\Delta \Phi$ of the potential jump is kept constant.  Same parameters as in Fig.~\ref{figeigen1}.}
\label{figeigendz}
\end{center}
\end{figure}

The boundary value problem is solved by integrating numerically the differential system~(\ref{dfdr1}-\ref{dsp}) from the boundary conditions (\ref{bc1}-\ref{Ssh}) at the shock to the lower boundary, using a  Newton-Raphson algorithm to find the discrete set of eigenfrequencies such that the lower boundary condition (\ref{bcleaking}) is satisfied.
Fig.~\ref{figeigen1} illustrates the typical irregular shape of the eigenspectrum of this type of flows. The eigenfrequencies that are plotted correspond to the most unstable harmonic for each transverse number $0\le n_x\le 7$.
The following set of parameters has been used: ${c_{\rm in}^2/c_{\rm out}^2}=0.75$, $L_x=4$, $\Deltaz_{\nabla}=0.1$, where lengthscales are expressed in units of $\Deltaz$. Eigenfrequencies are shown using a reference timescale $\tau_{\rm aac}$ defined as the sum of the advective and longitudinal acoustic timescales between the shock and the deceleration region:
\begin{eqnarray}
\tau_{\rm aac}\equiv {\Deltaz\over c_{\rm in}}\;\;{1\over \M_{\rm in}(1-\M_{\rm in})}.
\end{eqnarray}
Although the mode $n_x=1$ is the most unstable in Fig.~\ref{figeigen1}, this is not a strict rule. Fig.~\ref{figeigendz} shows that the order $n_x$ of the most unstable mode depends on the size of the coupling region. For example, the linear instability should be dominated by a mode $n_x=4$ if $\Deltaz_\nabla=0.04$, and by $n_x=7$ if $\Deltaz_\nabla=0.02$. 

According to Fig.~\ref{figeigendz}, low order modes seem to be favoured by the large size of the coupling region. This observation will be confirmed by physical arguments in Sect.~5. 
Note that the stability of the longitudinal mode $n_x=0$ is not an intrinsic feature of this toy model, but the consequence of tuning the size of the potential jump. In the limit $\Deltaz_\nabla\to 0$, the mode $n_x=0$ would become unstable if ${c_{\rm in}^2/ c_{\rm out}^2}\le0.7167$, but even then it would still be less unstable than transverse modes. Even for a weak potential jump such as ${c_{\rm in}^2/ c_{\rm out}^2}=0.9$, some modes are still unstable (\eg $n_x=18$) if the coupling region is sufficiently narrow. No unstable modes where found for ${c_{\rm in}^2/ c_{\rm out}^2}\ge0.95$.

\section{Method for the determination of the instability mechanism 
\label{Sect_method}}

\subsection{Cycles efficiencies ${\cal Q}$ and ${\cal R}$}

The determination of the eigenfrequencies in Fig.~\ref{figeigen1} is not sufficient to reveal the instability mechanism at work. Our toy model has been designed to be simple enough to allow for an unambiguous determination of the mechanism, using the same method as in Foglizzo (2002) for black hole accretion or Foglizzo et al. (2007) and YF08 for neutron star accretion. In these studies, this method was restricted to perturbations with a high enough frequency for the validity of the WKB approximation in the vicinity of the shock. This condition excluded the description of waves propagating horizontally along the shock. By contrast, the simplicity of the present set up allows for an exact calculation even at low frequency. It leads to an analytical formulation of the advective-acoustic instability in a compact approximation in the same spirit as Marble \& Candel (1977) or Foglizzo \& Tagger (2000) in 1D. The originality of the present formulation is that it includes transverse motions and vorticity perturbations, since our model is 2D. This toy model can thus describe the vortical-acoustic cycle as well as the entropic-acoustic cycle.\\
Our method consists in analyzing the interaction of plane waves (\ie with a real frequency) with both the shock and the region of deceleration. Perturbations with a given frequency $\omega_r$ and transverse wavenumber $k_x$ are locally projected onto pressure waves and advected waves (entropy/vorticity) (see Appendix~A). These different waves are independent from each other in each uniform part of the flow, but get coupled together both at the shock and in the region of flow gradients. This coupling is measured by two efficiencies ${\cal R}_{\rm sh}$ and ${\cal R}_{\nabla}$ of acoustic reflection, and two efficiencies ${\cal Q}_{\rm sh}$, and ${\cal Q}_{\nabla}$ of advective-acoustic coupling, illustrated in Fig.~\ref{figtoy}. Connecting these waves into two global cycles, one can estimate the global efficiency ${\cal Q}\equiv {\cal Q}_{\rm sh}\times {\cal Q}_{\nabla} $ of the advective-acoustic cycle, and the global efficiency ${\cal R}\equiv {\cal R}_{\rm sh}\times {\cal R}_{\nabla} $ of the purely acoustic cycle in order to address the question of the stability of each of these cycles.\\
As illustrated by Fig.~\ref{figtoy}, the coefficient ${\cal Q}_\nabla$ is the net effect of the advective-acoustic coupling associated to the region of deceleration, together with the advection from the shock to this region and the acoustic propagation from it up to the shock. Similarly, the coefficient ${\cal R}_\nabla$ includes the acoustic propagation between the shock and the region of deceleration, in addition to the acoustic reflection by the deceleration region.\\
Immediately after the shock, the perturbation of energy density is decomposed into acoustic ones propagating downward ($\delta f_{\rm sh}^+$), upward ($\delta f_{\rm sh}^-$) and advected ones ($\delta f_{\rm sh}^S$):
\begin{eqnarray}
\delta f_\sh&=&\delta f_\sh^S+\delta f_\sh^++\delta f_\sh^-.
\end{eqnarray}
The definitions of the coupling constants ${\cal R}_\sh$, ${\cal Q}_\sh$, ${\cal R}_\nabla$, ${\cal Q}_\nabla$ correspond to the following relations:
\begin{eqnarray}
\delta f_\sh^+&=&{\cal R}_\sh \delta f_\sh^-,\\
\delta f_\sh^S&=&{\cal Q}_\sh \delta f_\sh^-,\\
\delta f_\sh^-&=&{\cal Q}_\nabla \delta f_\sh^S+{\cal R}_\nabla \delta f_\sh^+.\label{decompsh}
\end{eqnarray}
Suppressing $\delta f_{\rm sh}^\pm$, $\delta f_{\rm sh}^S$ from these relations,
\begin{eqnarray}
{\cal Q}+{\cal R}=1,\label{bicycle}
\end{eqnarray}
with
\begin{eqnarray}
{\cal Q}&\equiv&{\cal Q}_\sh{\cal Q}_\nabla,\label{defq}\\
{\cal R}&\equiv&{\cal R}_\sh{\cal R}_\nabla.\label{defr}
\end{eqnarray}
The analytic calculation of the coupling efficiencies ${\cal Q}_{\rm sh}$, ${\cal R}_{\rm sh}$ at the shock was done approximately in Foglizzo et al. (2005), and the exact calculation is repeated below in Sect.~4. \\
The calculation of the coupling efficiencies ${\cal Q}_{\nabla}$, ${\cal R}_{\nabla}$ requires in principle an integration across the region of coupling, as was done in Foglizzo (2002). Taking advantage of the simplicity of the present flow, we show in Sect.~5 that an algebraic expression of the coupling efficiencies ${\cal Q}_{\nabla}$ and ${\cal R}_{\nabla}$ is possible in the compact approximation. \\
Anticipating on the results of Sect.~6, this calculation concludes that the purely acoustic cycle is always stable ($\ie |{\cal R}|\le1$), and that only the advective-acoustic cycle may be unstable (\ie $|{\cal Q}|>1$). This approach also proves that transverse perturbations are more unstable than longitudinal ones.

\subsection{Cycle of waves and growth rate: a new method to estimate the contribution of each cycle\label{sectaac}}

In Foglizzo (2002) and Foglizzo et al. (2007, hereafter FGSJ07), the efficiencies $|{\cal Q}|$, $|{\cal R}|$ of the cycles of plane waves have been computed for a {\it real} frequency $\omega=\omega_r$. Eq.~(\ref{bicycle}) invites us to evaluate the contribution of each cycle considered alone through the complex frequency $\omega^{\rm aac}$ solution of 
${\cal Q}(\omega)=1$ for the advective-acoustic cycle, or the complex frequency $\omega^{\rm pac}$ solution of ${\cal R}(\omega)=1$ for the purely acoustic cycle.
The eigenfrequencies $\omega^{\rm aac}$, $\omega^{\rm pac}$ associated in this manner to each cycle can be compared to the eigenfrequency $\omega$ computed directly in Sect.~2, in order to assess their relative importance in the instability mechanism. This method is used for the first time in Sect.~\ref{Sect_wcycle}.

\section{Efficiency of acoustic reflection and advective-acoustic coupling at the shock \label{secQRsh}}

\subsection{Acoustic reflection at a shock}

\begin{figure}
\begin{center}
\includegraphics[width=\columnwidth]{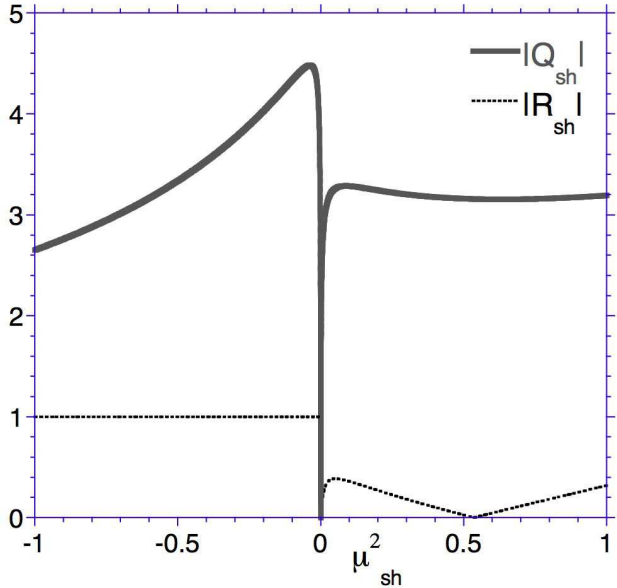}
\caption[]{Efficiencies $|{\cal Q}_\sh|$ and $|{\cal R}_\sh|$ of the advective-acoustic coupling and acoustic reflection, deduced from Eqs.~(\ref{Rsh}) and (\ref{eqQsh}). These efficiencies depend on the frequency of the perturbation only through the parameter $\mu_\sh^2$. }
\label{figQRsh}
\end{center}
\end{figure}

The efficiency of acoustic reflection has been computed by Foglizzo et al. (2005) 
(Eq.~(F10)):
\begin{eqnarray}
{\cal R}_\sh&\equiv&{\delta f^{+}_\sh\over \delta f^-_\sh}
={1+\mu_\sh\M_\sh\over1-\mu_\sh\M_\sh}\;\;{\delta p^{+}_\sh\over \delta p^-_\sh}
,\label{defRsh}\\
&=&-
{\mu_\sh^{2}- 
2\M_\sh\mu_\sh+\M_{1}^{-2}\over
\mu_\sh^{2}+  
2\M_\sh\mu_\sh+\M_{1}^{-2}}\;\;
{1+\mu_\sh\M_\sh\over1-\mu_\sh\M_\sh}.\label{Rsh}
\end{eqnarray}
The WKB approximation in Eq.~(\ref{fwkb}) of Appendix~B (or Eq.~(C1) in Foglizzo 2002) indicates that 
$\mu|\delta f^\pm|^2/vc$ is conserved to second order in an adiabatic flow. This guarantees that the quantity $|{\cal R}_\sh|^2$ can also be interpreted as a ratio of fluxes of acoustic energies.\\
From the expression of $k_z^\pm$ in Eq.~(\ref{ky}), acoustic waves are propagating in the $z$ direction if $\mu^2>0$ (i.e. $\omega>\omega_{\ev}$) and evanescent otherwise. Using Eq.~(\ref{Rsh}), one can prove that $|{\cal R}_\sh|<1$ for  propagating acoustic waves ($\mu_\sh^2>0$) and $|{\cal R}_\sh|=1$ for evanescent ones ($\mu_\sh^2<0$), as illustrated by the dotted line in Fig.~\ref{figQRsh}. This property precludes the possibility of an over-reflection, at least in this simple toy model.

\subsection{Advective-acoustic coupling at a shock}

The coupling coefficient obtained in the WKB approximation in Foglizzo et al. (2005) (Eq.~(F11-F12)) can be rewritten as an exact formulae in the present toy model since the flow is uniform immediately after the shock:
\begin{eqnarray}
{\cal Q}_\sh&\equiv&{\delta f^S_\sh\over \delta f^-_\sh}
={1\over1-\mu_\sh\M_\sh}\;\;{p_\sh\delta S_\sh\over\delta p_\sh^-}
,\\
&=&{2\over\M_\sh}\;\;{1-\M_\sh^2\over 1+\gamma\M_\sh^2}\left(1-{\M_\sh^2\over\M_1^2}\right)\nonumber\\
&&
\times
{\mu_\sh\over
(1-\mu_\sh\M_\sh)(\mu_\sh^2+2\mu_\sh\M_\sh+\M_1^{-2})},\label{eqQsh}
\end{eqnarray}
This formula is checked by numerical simulations in paper~II. Note that $|{\cal Q}_{\rm sh}|^2$ is not a ratio of energy fluxes (unlike $|{\cal R}_{\rm sh}|^2$).\\
Eq.~(\ref{eqQsh}) indicates that the efficiency of the advective-acoustic coupling vanishes in the limit of a weak shock: ${\cal Q}_\sh\propto(1-\M_\sh^2)$ for radial perturbations ($\mu_{\rm sh}=1$), and ${\cal Q}_\sh\propto(1-\M_\sh^2)^2$ for transverse ones ($\mu_{\rm sh}<1$). As seen on Fig.~\ref{figQRsh}, there are two optimal values of $\mu_\pm^2\propto\pm1/\M_1^2$ leading to a maximal production $|{\cal Q}_\sh|$ of advected perturbations. In the asymptotic limit of a strong shock, the maximum efficiencies are reached for acoustic waves propagating horizontally ($\omega\sim \omega_{\ev}^{\rm sh}$):
\begin{eqnarray}
|{\cal Q}_\sh|&\sim&{1\over\M_\sh^2}\;\;{1-\M_\sh^2\over 1+\gamma\M_\sh^2}.
\end{eqnarray}

\section{Computation of the acoustic feedback\label{secQfb}}

\subsection{Analytic expressions of ${\cal R}_\nabla$, ${\cal Q}_\nabla$ in the compact approximation}

\subsubsection{Range of validity}

The ``compact approximation" consists in treating the inhomogeneous layer as infinitely thin. This approximation was used by Marble \& Candle (1977) in the context of jet nozzles, in order to compute the acoustic feedback from advected entropy perturbations, in 1D. 
It is here extended to parallel flows in an external potential $\Phi$, subject to both entropy and vorticity perturbations. 
This approximation is valid for perturbations such that their vertical wavelength $2\pi/k_z$ is long compared to the lengthscale $\Deltaz_\nabla$ of the region of deceleration, on both sides of it. For acoustic waves, the criterion $k_z^\pm\Deltaz_\nabla\ll 2\pi$ can be approximated, using Eq.~(\ref{ky}) and $c_{\rm in}<c_{\rm out}$, by the following sufficient condition on the frequency $\omega$ of the wave:
\begin{eqnarray}
\omega\ll \left\lbrack k_x^2 + \left({2\pi\over\Deltaz_\nabla}\right)^2\right\rbrack^{1\over2}c_{\rm in}.\label{compacc}
\end{eqnarray}
For advected waves ($k_z=\omega/v$), the frequency threshold for the compact approximation can be related to the time $\tau_\nabla$ of advection through the inhomogeneous region:
\begin{equation}
{2\pi\over\omega}\gg\tau_\nabla\equiv
\int_{z_\nabla-{\Deltaz_\nabla\over2}}^{z_\nabla+{\Deltaz_\nabla\over2}} {\dd z\over |v|}.\label{compact}
\end{equation}
The condition (\ref{compact}) for advected waves is more severe than (\ref{compacc}) because the flow is subsonic. The coefficients of acoustic refraction, and advective-acoustic coupling can be expressed through simple algebraic formulae based on the conservation of entropy, mass flux and energy density ($\delta f$, $\delta h$ and $\delta S$) through the potential jump.

\subsubsection{Coupling efficiencies ${\cal R}_{\nabla}$ and ${\cal Q}_{\nabla}$\label{Sectnabla}}

\begin{figure}
\begin{center}
\includegraphics[width=\columnwidth]{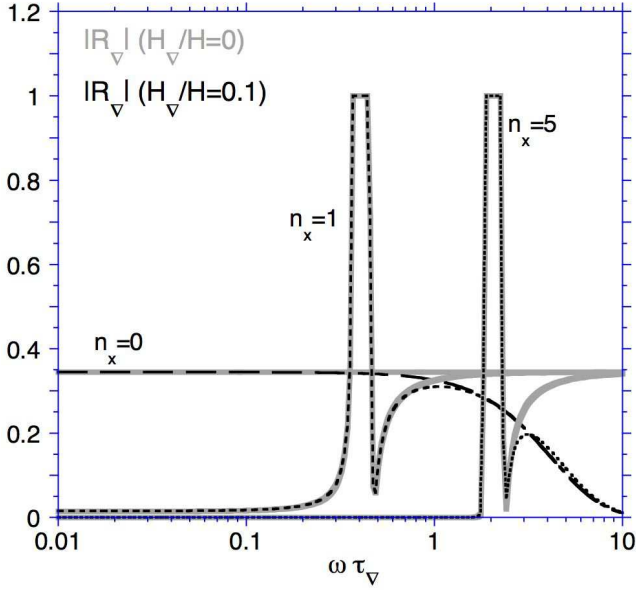}
\caption[]{Refraction efficiency $|{\cal R}_\nabla|$ for longitudinal and transverse perturbations, in a flow where $\Deltaz_\nabla/\Deltaz=0.1$. The different curves are labelled with the value of the transverse wavenumber $n_x=0,1,5$. The thick grey lines are the analytical estimates deduced from the compact approximation, in Eq.~(\ref{calR}). The range of frequencies corresponding to total reflection coincides with $\lbrack \omega_{\ev}^{\rm in},\omega_{\ev}^{\rm out}\rbrack$ (Eq.~(\ref{omegaz}))}
\label{figRk}
\end{center}
\end{figure}
\begin{figure}
\begin{center}
\includegraphics[width=\columnwidth]{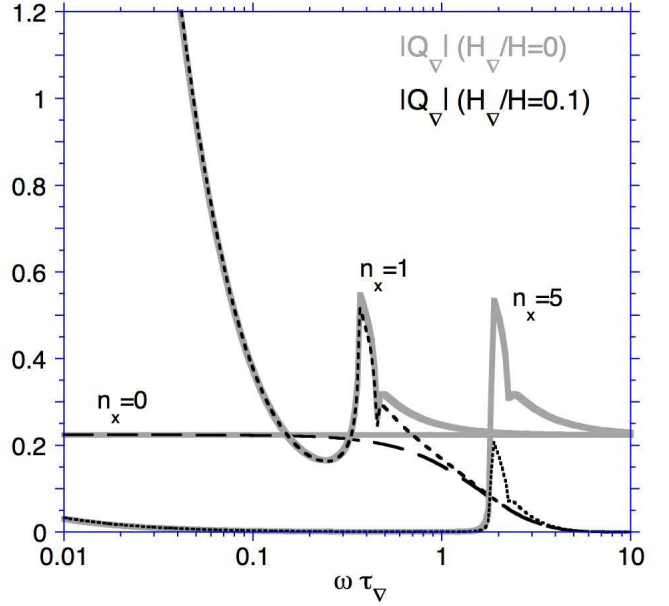}
\caption[]{Efficiency $|{\cal Q}_\nabla|$ of the advective-acoustic coupling in the same flow as in
Fig.~\ref{figRk}. The thick grey lines corresponds to Eq.~(\ref{calQS}). Kinks in the curves correspond to 
the particular frequencies $\omega_{\ev}^{\rm in}$ and $\omega_{\ev}^{\rm out}$. There is no frequency cut-off in the compact approximation.}
\label{figQSk}
\end{center}
\end{figure}

In Appendix~C, the following expressions are deduced from the conservation laws : 
\begin{eqnarray}
{\cal R}_{\nabla}&=&{\mu_{\rm in}\M_{\rm out}c_{\rm out}^2
-\mu_{\rm out}\M_{\rm in}c_{\rm in}^2
\over \mu_{\rm in}\M_{\rm out}c_{\rm out}^2+\mu_{\rm out}\M_{\rm in}c_{\rm in}^2}
\e^{i\omega\tau_{\cal R}},\label{calR}\\
{\cal Q}_{\nabla}&=&{\M_{\rm out}+\mu_{\rm out}\over 1+\mu_{\rm out}\M_{\rm out}}
\;\;{\e^{i\omega\tau_{\cal Q}}\over 
\mu_{\rm out}{c^2_{\rm in}\over c^2_{\rm out}}+
\mu_{\rm in}{\M_{\rm out}\over\M_{\rm in}}}
\nonumber\\
&&\times\left\lbrack
1-{c^2_{\rm in}\over c^2_{\rm out}}
+{k_x^2c_{\rm in}^2\over \omega^2}(\M_{\rm in}^2-\M_{\rm out}^2)
\right\rbrack,\label{calQS}
\end{eqnarray}
where $\tau_{\cal Q}$ and $\tau_{\cal R}$ are defined by:
\begin{eqnarray}
\tau_{\cal Q}&\equiv&\tau_{\rm aac}{1+\mu_{\rm in}\M_{\rm in}\over1+\M_{\rm in}},\label{tauQ1}\\
\tau_{\cal R}&\equiv&{\Deltaz\over c_{\rm in}}\;\;{2\mu_{\rm in}\over1-\M_{\rm in}^2}.\label{tauR1}
\end{eqnarray}
As expected, these coupling efficiencies vanish in the limit of a uniform flow ($c_{\rm out}=c_{\rm in}$ and Eq.~(\ref{Moutin})). \\
The exponential functions in Eqs.~(\ref{calR}-\ref{calQS}) are associated to the vertical structure of advected and acoustic waves. In the low frequency limit ($\omega\ll \omega_{\ev}^{\rm in}$), the evanescent character of acoustic waves is responsible for the following damping effect:
\begin{eqnarray}
\left|\e^{i\omega\tau_{\cal R}}\right|&\sim&\exp\left\lbrack{-2k_x\Deltaz\over(1-\M_{\rm in}^2)^{1\over2}}
\right\rbrack,\label{evaneR}\\
\left|\e^{i\omega\tau_{\cal Q}}\right|&\sim&\exp\left\lbrack{-k_x\Deltaz\over(1-\M_{\rm in}^2)^{1\over2}}\right\rbrack.\label{evaneQ}
\end{eqnarray}
This acoustic evanescence is a strong stabilizing factor for any cycle based on high degree perturbations ($n_x\gg L_x/2\pi\Deltaz$).

\subsection{Comparison with the exact calculation: the frequency cut-off $\omega_\nabla$\label{secQRnabla}}

\begin{figure}
\begin{center}
\includegraphics[width=\columnwidth]{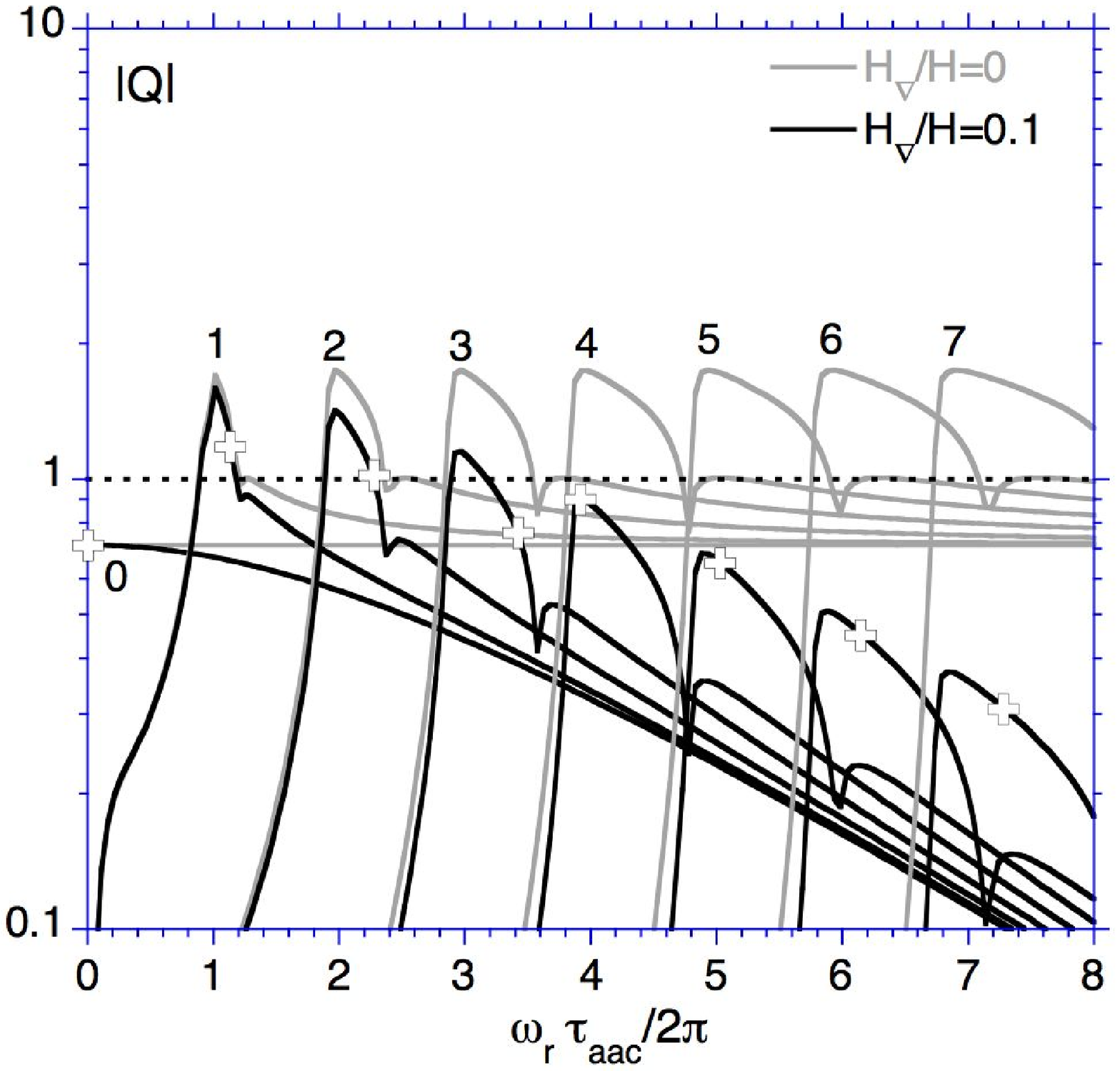}
\includegraphics[width=\columnwidth]{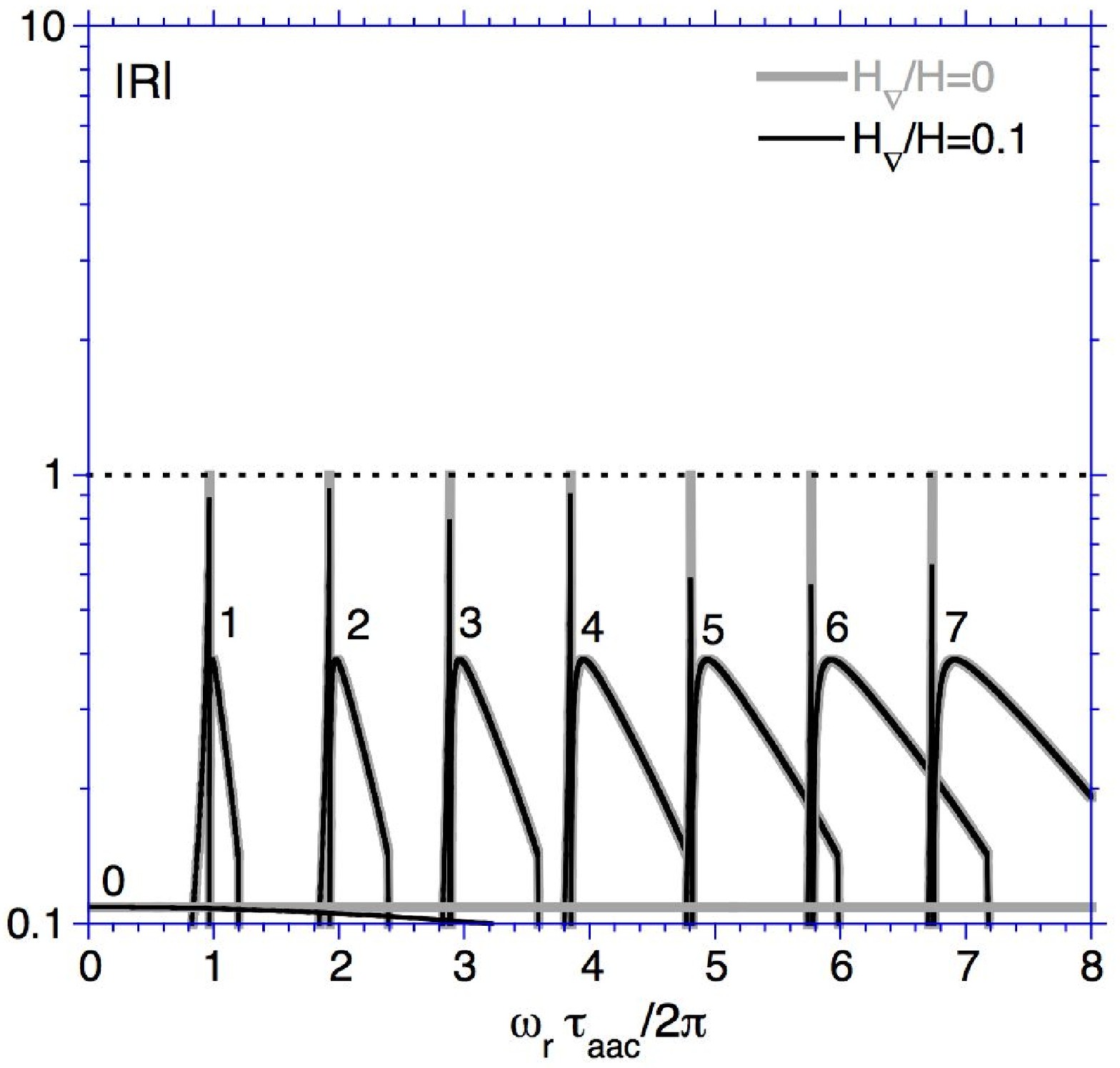}
\caption[]{Efficiencies $|{\cal Q}|$ and $|{\cal R}|$ of the advective-acoustic cycle in a flow where $\Deltaz_\nabla/\Deltaz=0.1$ (black lines). The different curves are labelled with the value of the transverse wavenumber $0\le n_x\le7$. In the upper plot, the crosses indicate the discret set of oscillation frequencies corresponding to the solution of $Q=1$ (Fig.~\ref{figspectrQR}). The grey lines are the analytical estimates deduced from the compact approximation. For comparison with Fig.~\ref{figQSk}, $\tau_{\rm aac}/\tau_\nabla\sim 16.52$.}
\label{figQtotal}
\end{center}
\end{figure}

Figures~\ref{figRk} and~\ref{figQSk} show the efficiencies $|{\cal Q}_\nabla|$, $|{\cal R}_\nabla|$ computed in the compact approximation (thick grey lines) and the exact calculation with  $\Deltaz_\nabla/\Deltaz=0.1$ (thin black lines). The numerical method to compute ${\cal Q}_\nabla$, ${\cal R}_\nabla$ is based on three numerical integration from the shock to the lower boundary, using different boundary conditions at the shock, and two appropriate linear combinations of these three solutions such that the lower boundary condition is fulfilled, as explained in Appendix~D of FGSJ07. ${\cal Q}_\nabla$ can also be computed as an explicit integral over the region of flow gradients, using Green functions as explained in the Appendix~D of the present paper. Schematically, the global advective-acoustic coupling ${\cal Q}_\nabla$ is described by a convolution between the radial profile of acoustic waves $\delta p_0/p$, the radial profile of advected waves $\e^{\int{i\omega\over v}\dd z}$, and the local emissivity $\p b_\nabla/\p z$ associated to the gradients of the flow:
\begin{eqnarray}
{\cal Q}_\nabla&=&
\int_{\rm bc}^{\rm sh}b_0{\delta p_0\over p} \e^{\int_{\rm sh}{i\omega\over v}\dd z}{\p b_\nabla\over\p z}\dd z,
\label{qnabgrad}
\end{eqnarray}
where 
\begin{eqnarray} 
b_0&\equiv&{1\over2 }
\left(1+{k_x^2v_{\rm sh}^2\over \omega^{2}}\right)
\left(1-{\cal R}_\nabla-{1+{\cal R}_\nabla\over\mu_{\rm sh}\M_{\rm sh}}\right)
\nonumber\\
&&{1-\M^2\over1-\M^2_{\rm sh}} {\M^2_{\rm sh}\over\M^2}
\left({\delta p_0\over p}\right)_{\rm sh}^{-1}
\e^{-\int_{\rm sh}{i\omega\over c}{2\M\over1-\M^2}\dd z},\\
b_\nabla&\equiv&
{i\omega\over c_{\rm sh}^2}\;\;
{i\omega-2v{\p\log\M\over\p z}\over
k_x^2\M^2+{\omega^2\over c^2}-v\M^2{\p\over\p z}\;{i\omega\over v^2} }
.\label{gradients}
\end{eqnarray}
In comparison with Eq.~(25) of Foglizzo (2001), the present formulation stresses the local character of the acoustic emissivity associated to the flow gradients, described by Eq.~(\ref{gradients}). \\
Figures~\ref{figRk} and~\ref{figQSk} show that the analytical compact approximation is excellent for $\omega\tau_{\nabla}\ll1$, and works quite well even for values just below one. 
As expected, $\tau_{\nabla}^{-1}$ defines a cut-off frequency above which the compact approximation 
ceases to be valid. From Fig.~\ref{figQSk}, one can estimate the cut-off associated to the advective-acoustic coupling:
\begin{eqnarray}
\omega_\nabla\sim {1\over\tau_{\nabla}}.\label{omeganabla}
\end{eqnarray}
Comparing Figs.~\ref{figRk} to Fig.~\ref{figQSk}, the frequency cut-off associated to ${\cal Q}$ is smaller than the one associated to ${\cal R}$, as expected from Eqs.~(\ref{compacc}) and (\ref{compact}). The different nature of these cut-off frequencies can be understood owing to Eq.~(\ref{qnabgrad}): the efficiency ${\cal Q}_\nabla$ decreases not only if the wavelength of the perturbations is short compared to the lengthscale of the flow gradients (i.e. the term $\p b_\nabla/\p z$), but also if the convolution of the advected and acoustic phase structures leads to a phase averaging. Also remarkable in Fig.~\ref{figRk} is the fact that acoustic waves propagating horizontally ($\omega\sim \omega_\ev^{\rm in}$) are always well described by the compact approximation, with $|{\cal R}_\nabla|=1$ (this is also visible in the bottom plot of Fig.~\ref{figQtotal}). Altogether, Figs.~\ref{figRk} and \ref{figQSk} indicate that the compact approximation may be used as an upper bound for the efficiency of the acoustic feedback. \\
Acoustic reflection is always damped ($|{\cal R}_\nabla|\le1$). Total reflection without damping ($|{\cal R}_\nabla|=1$) occurs for transverse perturbations in the range of frequencies $\lbrack \omega_{\ev}^{\rm in},\omega_{\ev}^{\rm out}\rbrack$ where both $\mu_\inn^2>0$ and $\mu_\out^2<0$, \ie when acoustic waves propagate above the inhomogeneous region and are evanescent below it.\\
A striking feature on Fig.~\ref{figQSk} is the divergence $|{\cal Q}_\nabla|\propto \omega^{-1}$ of the coupling efficiency at low frequency for transverse perturbations (Eq.~(\ref{calQS})). This divergence however is not sufficient to overcome the weak efficiency of production of advected perturbations at the shock, $|{\cal Q}_\sh|\propto \omega^{2}$ (Eq.~(\ref{eqQsh})). The frequency dependence of the global efficiency $|{\cal Q}|=|{\cal Q}_{\rm sh}{\cal Q}_\nabla|$ is further studied in the next section (Fig.~\ref{figQtotal}).

\section{Global cycle\label{secglobal}}

\begin{figure}
\begin{center}
\includegraphics[width=\columnwidth]{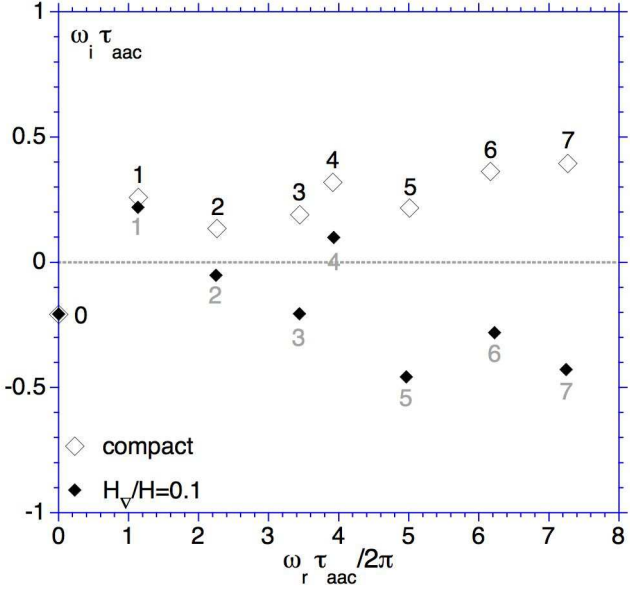}
\caption[]{Eigenfrequencies of the most unstable modes for $0\le n_x\le 7$ in a flow where $\Deltaz_{\nabla}=0$ (empty diamonds). The value $n_x$ of the corresponding wavenumber is indicated next to each eigenfrequency.
Increasing the size $\Deltaz_{\nabla}$ of the deceleration region up to $\Deltaz_{\nabla}/\Deltaz=0.1$ is stabilizing (filled diamonds).}
\label{figeigencompact}
\end{center}
\end{figure}

\subsection{Compact approximation of ${\cal Q}$ and ${\cal R}$}

The coefficients ${\cal Q}$ and ${\cal R}$ defined in Eqs.~(\ref{defq}-\ref{defr}) can be expressed analytically by combining Eqs.~(\ref{eqQsh}) and (\ref{calQS}) for ${\cal Q}$, and Eqs.~(\ref{Rsh}) and (\ref{calR}) for ${\cal R}$:
\begin{eqnarray}
{\cal Q}&=&{\cal Q}_{0}\e^{i\omega\tau_{\cal Q}},\\
{\cal R}&=&{\cal R}_{0}\e^{i\omega\tau_{\cal R}},\label{rr0}\\
{\cal Q}_{0}&\equiv&{4\over\gamma+1}\;\;{\mu_{\rm in}\over\M_{\rm in}}\;\;
{1-{c^2_{\rm in}\over c^2_{\rm out}}
+{k_x^2c_{\rm in}^2\over\omega^2}(\M_{\rm in}^2-\M_{\rm out}^2)
\over
(1-\mu_{\rm in}\M_{\rm in})
(\mu_{\rm in}^2+2\mu_{\rm in}\M_{\rm in}+\M_1^{-2})}\nonumber\\
&&\times{\M_{\rm out}+\mu_{\rm out}\over 1+\mu_{\rm out}\M_{\rm out}}
\;\;{(1-\M_{\rm in}^2)(1-\M_1^{-2})\over 
\mu_{\rm out}{c^2_{\rm in}\over c^2_{\rm out}}+
\mu_{\rm in}{\M_{\rm out}\over\M_{\rm in}}},\label{fullQ}\\
{\cal R}_{0}&\equiv&
-\left({1+\mu_{\rm in}\M_{\rm in}\over1-\mu_{\rm in}\M_{\rm in}}\right)
{\mu_{\rm in}^2-2\mu_{\rm in}\M_{\rm in}+\M_1^{-2}\over
\mu_{\rm in}^2+2\mu_{\rm in}\M_{\rm in}+\M_1^{-2}}\nonumber\\
&&\times{\mu_{\rm in}\M_{\rm out}c_{\rm out}^2
-\mu_{\rm out}\M_{\rm in}c_{\rm in}^2
\over \mu_{\rm in}\M_{\rm out}c_{\rm out}^2+\mu_{\rm out}\M_{\rm in}c_{\rm in}^2}.\label{fullR}
\end{eqnarray}
The cycles efficiencies $|{\cal Q}|$ and $|{\cal R}|$ are shown as functions of the real frequency $\omega_r$ in Fig.~\ref{figQtotal}, both in the compact approximation and for $\Deltaz_\nabla/\Deltaz=0.1$, for different values of the transverse wavenumber $n_x$. 
The stability of the purely acoustic cycle is confirmed by the bottom plot in Fig.~\ref{figQtotal}. The spikes correspond to the frequencies $\omega_{\ev}^{\rm in}$, for which $|{\cal R}|=1$ according to Eqs.~(\ref{defr}), (\ref{rr0}) and (\ref{fullR}).\\
The stabilization of the advective-acoustic cycle above the cut-off frequency is clearly demonstrated by the upper plot in Fig.~\ref{figQtotal}. The maximum coupling efficiency is always reached for $\omega\sim\omega_{\ev}^{\rm in}$, as expected from Sects.~4 and 5 since $\omega_{\ev}^{\rm in}=\omega_{\ev}^{\rm sh}$ in our toy model. If the shock is strong and $\M_\out\ll1$, we deduce from Eq.~(\ref{fullQ}) that the
maximum efficiency is
\begin{eqnarray}
{\cal Q}_{\rm max}&\sim&{2\over\gamma+1}\;\;{1\over\M_\inn^2}\;\;{c^2_{\rm out}\over c^2_{\rm in}}\gg1.
\end{eqnarray}
In our toy model with a strong shock, the maximum efficiency ${\cal Q}_{\rm max}$ is larger in 2D than in 1D by a factor which depends only on $\M_{\rm in}$: 
\begin{eqnarray}
{{\cal Q}_{\rm max}(2D)\over {\cal Q}_{\rm max}(1D)}
={1+2\M_{\rm in}\over2\M_{\rm in}(1+\M_{\rm in})}>1.\label{2D1D}
\end{eqnarray}
This factor is equal to $1.8$ for $\gamma=4/3$.\\
According to Fig.~\ref{figQtotal}, the advective-acoustic cycle alone is necessarily stable ($|{\cal Q}_{\rm max}|<1$) for $n_x\ge4$ if $\Deltaz_\nabla/\Deltaz=0.1$, and can be unstable for $n_x\le3$. Interestingly, Fig.~\ref{figeigen1} indicates that the mode $n_x=2$ is stable and the mode $n_x=4$ is unstable, whereas the upper plot of Fig.~\ref{figQtotal} shows that $|{\cal Q}|>1$ is possible for the mode $n_x=2$, and that $|{\cal Q}|<1$ for $n_x=4$. This apparent paradox is clarified below by taking into account two factors: 
\par (i) the phase closure relation (\ref{bicycle}) selects a discrete set of oscillation frequencies, which do not necessarily allow for the maximum coupling efficiency $|{\cal Q}|_{\rm max}$ to be reached,
\par (ii) the additional influence of the purely acoustic cycle can be constructive or destructive, and can be crucial close to marginal stability.\\

The closure relation (\ref{bicycle}) takes the following particular form in the compact approximation:
\begin{eqnarray}
{\cal Q}_{0}\e^{i\omega\tau_{\cal Q}}+{\cal R}_{0}\e^{i\omega\tau_{\cal R}}=1,\label{bicycle_c}
\end{eqnarray}
This explicit analytical equation is a 2D generalization of the equation obtained in 1D by Foglizzo \& Tagger (2000). The corresponding eigenmodes, easily obtained numerically, are shown as empty diamonds in Fig.~\ref{figeigencompact}. Comparing the compact eigenspectrum (filled diamonds) to the one obtained for $\Deltaz_\nabla/\Deltaz=0.1$ (empty diamonds) shows again the stabilization above the cut-off frequency. 

As a consistency check, one can verify for $1\le n_x\le 7$ that the growth rates shown in Fig.~\ref{figeigendz}, in the limit $\Deltaz_\nabla/\Deltaz\ll1$, do converge towards the growth rates represented as empty diamonds in Fig.~\ref{figeigencompact}. The former are obtained by numerical integration of the boundary value problem, whereas the latter are directly deduced from the explicit Eqs.~(\ref{fullQ},\ref{fullR},\ref{bicycle_c}).

\subsection{Calculation of $\omega^{\rm aac}$, $\omega^{\rm pac}$\label{Sect_wcycle}}

\begin{figure}
\begin{center}
\includegraphics[width=\columnwidth]{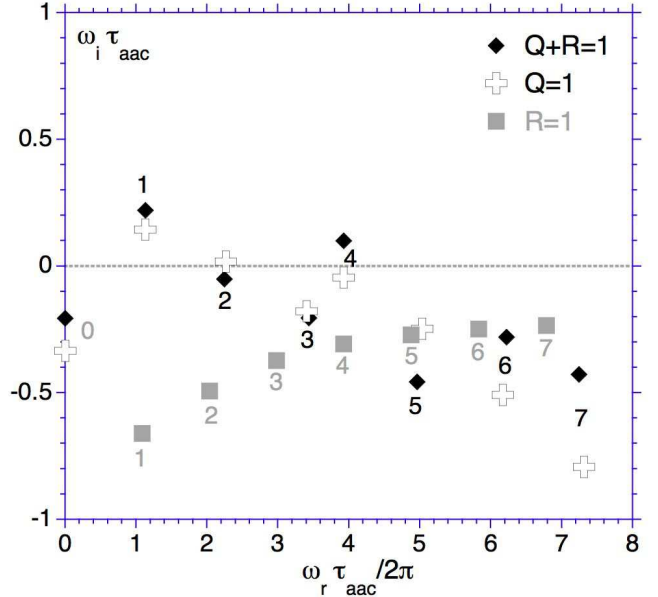}
\caption[]{For each value of $n_x=0$ to $7$, comparison of the most unstable modes associated to the advective-acoustic cycle alone (filled grey squares), the purely acoustic cycle alone (black crosses) and the full problem (filled black diamonds). The contribution of the purely acoustic cycle can either be constructive or destructive.}
\label{figspectrQR}
\end{center}
\end{figure}

As explained in Sect.~\ref{sectaac}, the complex eigenspectrum associated to each cycle considered alone is defined by the equations ${\cal Q}=1$ for the advective-acoustic cycle, and ${\cal R}=1$ for the purely acoustic cycle. 
These eigenspectra are compared to the solution of the full problem (${\cal Q}+{\cal R}=1$) in Fig.~\ref{figspectrQR}, where we have selected the most unstable mode for each degree $0\le n_x\le 7$. The eigenspectrum of the purely acoustic cycle is stable as expected. The eigenmodes of the advective-acoustic cycle match remarkably well the eigenmodes of the full problem from the point of view of the oscillation frequencies, and also the growth rates for $0\le n_x\le4$. The difference of growth rate between the solutions of ${\cal Q}+{\cal R}=1$ and ${\cal Q}=1$ is attributed to the contribution of the purely acoustic cycle, which is clearly constructive for the modes $n_x=0,1,4,6,7$ and destructive for the modes $n_x=2,5$. As already noted by Foglizzo \& Tagger (2000) and Foglizzo (2002), the contribution of the purely acoustic cycle is most important close to marginal stability: Fig.~\ref{figspectrQR} shows that the mode $n_x=2$ would be slightly unstable with the advective-acoustic cycle alone, and is stable because of the destructive interference with the acoustic cycle. Conversely, the mode $n_x=4$ would be stable with the advective-acoustic cycle alone, and is unstable because of the constructive interference with the acoustic cycle.\\
Unlike the method based on ${\cal Q}(\omega_r)$ (e.g. Foglizzo 2002, FGSJ07), the present method is not restricted to $\omega_i\ll\omega_r$.

\subsection{Analytical estimate of the maximum growth rate $\omega_i^{\rm aac}$ and sensitivity to geometrical factors}

\begin{figure}
\begin{center}
\includegraphics[width=\columnwidth]{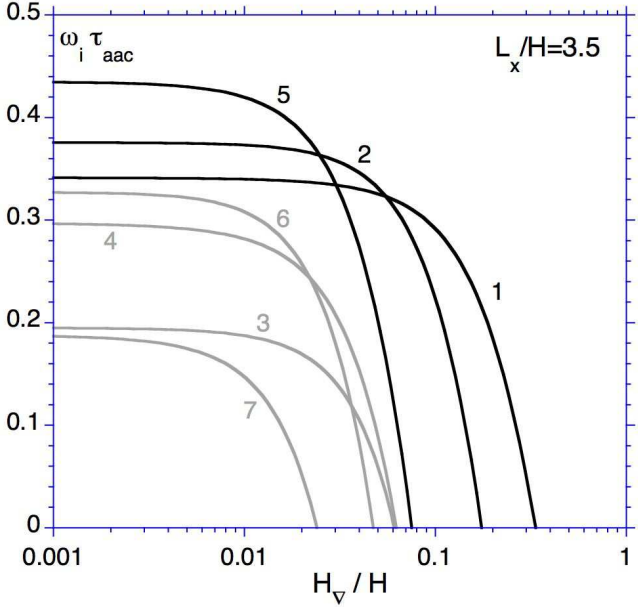}
\caption[]{Same as Fig.~\ref{figeigendz}, with $L_x/\Deltaz=3.5$ instead of $L_x/\Deltaz=4$. This small change in the aspect ratio of the flow changed the degree $n_x$ of the dominant mode: the modes $n_x=1$, $2$ and $5$ are now the dominant ones depending on the size of the region of coupling.}
\label{figeigendz2}
\end{center}
\end{figure}

Writing $\omega^{\rm aac}=\omega^{\rm aac}_r+i\omega^{\rm aac}_i$,  the growth rate $\omega^{\rm aac}_i$ of the advective-acoustic cycle alone can be estimated by a first order expansion of ${\cal Q}_0\exp(\omega^{\rm aac}\tau_{\cal Q})=1$ in the compact approximation:
\begin{eqnarray}
\omega^{\rm aac}_i&\sim&{1+\M_{\rm in}\over \mu_r+\M_{\rm in}}\;\;{\mu_r\over\tau_{\rm aac}}\log|{\cal Q}_0(\omega^{\rm aac}_r)|,\label{imomegaac}\\
\mu_r^2&\equiv&1-\left({\omega_{\ev}^{\rm in}\over \omega^{\rm aac}_r}\right)^2.
\end{eqnarray}
This equation is valid when $\omega^{\rm aac}_i\ll\omega^{\rm aac}_r$ and $\mu_r^2>0$.
Equation~(\ref{imomegaac}) illustrates the competing effects introduced by transverse perturbations. On the one hand, 2D perturbations allow for a higher coupling efficiency ${\cal Q}_0$ (Fig.~\ref{figQtotal} and Eq.~\ref{2D1D}), but on the other hand the acoustic timescale for the feedback is longer for transverse waves (factor $\mu_r$ in Eq.~(\ref{imomegaac})). Whether the optimum balance between these two effects is reached or not depends on geometrical factors such as the aspect ratio $L_x/\Deltaz$ of the toy model. Indeed, the discrete set of oscillation frequencies selected by the closure relation is sensitive to $L_x$ through $k_x=2\pi n_x/L_x$. Besides, the constructive or destructive contribution of the acoustic cycle is sensitive to $\Deltaz$ through the ratio of the acoustic and advective-acoustic timescales.
A comparison between Fig.~\ref{figeigendz} and Fig.~\ref{figeigendz2} illustrates the fact that a small change in the aspect ratio of the unstable cavity, from $L_x/\Deltaz=4$ to $L_x/\Deltaz=3.5$, is sufficient to favour the modes $n_x=1,2,5$ rather than $n_x=1,4,7$.

\subsection{A low frequency, low order instability if the coupling region is wide}

The frequency cut-off $\omega_\nabla$ associated in Sect.~\ref{secQRnabla} to the finite size of the coupling region is responsible for the stabilization of high frequency modes. Besides, the evanescent pressure feedback associated to high $n_x$ transverse modes (Eq.~\ref{evaneQ}) is inefficient at perturbing the shock. These two arguments explain why the advective-acoustic instability favours low frequency, low order modes if the size of the coupling region is large, as seen on Figs.~\ref{figeigen1} and \ref{figeigendz}.\\
More precisely, the evanescence length $\lbrack{\rm Im}(k_z^-)\rbrack^{-1}$ of the pressure perturbation can be compared to the distance $H$ to the shock. Using Eq.~(\ref{ky}), the conditions Im$(k_z^-H)<1$ and $\omega<\omega_\nabla$ can be translated into a condition on the horizontal wavenumber $k_x$:
\begin{eqnarray}
k_x^2<{\omega_\nabla^2\over c_{\rm in}^2(1-\M_{\rm in}^2)}+{1-\M_{\rm in}^2\over H^2}.
\end{eqnarray}
Neglecting $\M_{\rm in}^2\ll1$ for the sake of simplicity, and using Eq.~(\ref{omeganabla}), the range of frequencies and wavenumbers which can produce an unstable cycle most efficiently are summarized as follows:
\begin{eqnarray}
\omega_{\ev}^{\rm in}&\le&\omega< {1\over\tau_\nabla},\\
n_x^2 &<& \left({L_x\over 2\pi c_{\rm in}\tau_\nabla}\right)^2
+\left({L_x\over 2\pi \Deltaz}\right)^2.\label{nxmax}
\end{eqnarray}
The r.h.s in Eq.~(\ref{nxmax}) depends both on the velocity profile in the deceleration region (phase averaging), and on the shock distance (acoustic evanescence). The first term is dominant in the compact approximation.

\section{Relation to the instability of the standing shock during core collapse}

A direct extrapolation of Eq.~(\ref{nxmax}) to a stalled shock of radius $R_{\rm sh}$ in spherical geometry would favour unstable modes with a low azimuthal degree $l$: 
\begin{eqnarray}
l^2 &<& \left({R_{\rm sh}\over c_{\rm in}\tau_\nabla}\right)^2
+\left({R_{\rm sh}\over \Deltaz}\right)^2.\label{lxmax}
\end{eqnarray}
The deceleration region responsible for the acoustic feedback has been identified in the numerical simulations of Scheck \etal (2008), from which we estimate, $80$ms after bounce in their model W00F: $R_{\rm sh}\sim 110$km, $\Deltaz\sim 50$km, $R_{\rm sh}/c_{\rm in}\sim 5$ms, $\tau_\nabla\sim 15$ms and $\Deltaz_\nabla/\Deltaz\sim 0.4$. Using these numbers in Eq.~(\ref{lxmax}) we deduce 
$l\le 2.2$. Despite the simplicity of our approximation, this estimate is very close to the degree $l=1,2$ of SASI observed in Scheck \etal (2008), as well as in most numerical simulations of the stationary shock involved in the core-collapse problem.
These simulations cover a broad range of physical ingredients and approximations, from stationary flows with a simplified cooling function (Blondin \& Mezzacappa 2006, FGSJ07) to the most detailed modeling of neutrino transport  and advanced equations of state in the core of a massive progenitor (e.g. Burrows et al. 2007, Marek \& Janka 2007). SASI has also been observed in the pioneering 3D simulations of Blondin \& Mezzacappa (2007) and Iwakami et al. (2008). The efficiencies ${\cal Q}$, ${\cal R}$ of the cycles have been computed by FGSJ07 in the context of a simplified stationary flow, proving that the advective-acoustic mechanism is responsible for SASI, at least when the shock is far enough to allow for the WKB approximation. By taking a step further in the simplification of the flow, the present toy model enabled us to better understand some generic properties of this instability. A frequency cut-off is associated to the timescale of advection through the flow gradients. Among the low frequency perturbations below this cut-off, the most unstable ones are those which approach the maximum advective-acoustic coupling efficiency ${\cal Q}$ with a moderate lengthening of the acoustic feedback timescale. Close to marginal stability, constructive interference with the purely acoustic cycle can be essential for the instability. In such cases however, its consequences may be negligible due to a lack of time for significant growth.\\
By considering a 2D planar flow, our toy model neglected the large scale gradients that would exist in a spherical flow, all the way between the shock and the neutron star. These large scale gradients are expected to contribute the adiabatic coupling of advected and acoustic perturbations at low frequency. Neglecting these gradients in the toy model was important in order to characterize analytically acoustic waves in the vicinity of the shock, even at low frequency (Eqs.~(\ref{deffpm}-\ref{defhpm})). Another important simplification in our toy model was the hypothesis of an adiabatic region of deceleration, which allowed us to identify the phase averaging effect responsible for the inefficiency of the acoustic feedback, above the frequency cut-off, through Eq.~(\ref{qnabgrad}).\\
Of course, the acoustic feedback produced by advected perturbations is expected to depend quantitatively on whether the deceleration is produced by a cooling layer or an adiabatic gravity step. 
For example, the mode $l=0$ would be necessarily stable in an idealized isothermal flow, because the only advected perturbations which can build a cycle in an isothermal flow are non radial vorticity perturbations ($l\ge1$). The relative contributions of entropy and vorticity perturbations to the acoustic feedback are thus expected to differ significantly from our toy model when non adiabatic processes are involved.
We believe however that our description of the cycles, and the existence of a cut-off frequency associated to the advection time through the deceleration layer, are general enough to be relevant to the framework of stalled accretion shock during core collapse. This toy model allows us to propose answers to the questions: why SASI is a low frequency, low degree $l$ instability, and why the eigenspectrum is irregular and sensitive to geometrical factors. 

\section{Conclusion\label{Sect_conc}}

A simple toy model has been analyzed in order to better understand the advective-acoustic cycle in a decelerated flow. This comprehensive study has reached the following conclusions:
\par -The instability mechanism is based on the amplification of the advective-acoustic cycle.
\par -The purely acoustic cycle is always stable in this flow.
\par -Optimal transverse perturbations are more unstable than longitudinal ones.
\par -The finite size of the coupling region is responsible for a frequency cut-off which favours low frequency perturbations.
\par -This frequency cut-off disfavours high order perturbations, which would lead to an inefficient evanescent pressure feedback: the most unstable modes are expected to be low frequency and low order.
\par -The maximum efficiency $|{\cal Q}|$ of the advective-acoustic cycle is governed by a compromise between 
the efficiency of the advective-acoustic coupling, which favours transverse perturbations, and a short cycle time which disfavours too horizontal acoustic propagation. 
\par -The eigenspectrum is irregular and sensitive to geometrical factors due to the phase condition which selects a discrete set of frequencies, and the interferences with the acoustic cycle.
\par -A new method has been described to measure the eigenspectrum associated to each cycle considered alone.
\par -This toy model can be used as a benchmark test for numerical simulations, in order to check for undesired numerical artifacts. The important roles played by the shock, the advection of perturbations and the propagation of acoustic waves make this toy model particularly challenging for numerical simulations (paper~II).
\par -The comparison with the properties of SASI suggests that our toy model captures some fundamental features, explaining in particular why SASI is a low frequency, low degree instability.\\
A good understanding of the mechanism behind SASI should help perform accurate numerical simulations of core collapse explosions, with particular attention to the mesh size and the boundary conditions.

\acknowledgements
It is a pleasure to acknowledge useful discussions with F. Masset, J. Guilet, T. Yamasaki, J. Sato, M. Liebend\" orfer and S. Fromang, and the constructive comments of an anomymous referee. This work benefited from the ANR grant associated to the ``Vortexplosion" project ANR-06-JCJC-0119. The author is thankful to the Kavli Institute for Theoretical Physics (KITP) for its stimulating program ÔÔThe Supernovae-GRB Connection,ÕÕ supported in part by the National Science Foundation under grant PHY 99-07949. 

\appendix

\section{Decomposition of perturbations in a uniform flow}

For completeness we repeat here the relationship between $\delta f$, $\delta h$ and $\delta v_z$, $\delta \rho$, $\delta p$ corresponding to Eqs.~(A1-A4) in FGSJ07:
\begin{eqnarray}
{\delta v_z\over v}&=&{1\over 1-\M^2}\left(\delta h+\delta S-{\delta f\over c^2}\right),\label{dvv}\\
{\delta \rho\over\rho}&=&{1\over 1-\M^2}\left(-\M^2\delta h-\delta S+{\delta f\over c^2}\right),\label{drho}\\
{\delta p\over \gamma p}&=&{1\over 1-\M^2}\left\lbrace
-\M^2\delta h-\left\lbrack 1+(\gamma-1)\M^2\right\rbrack{\delta S\over\gamma}
+{\delta f\over c^2}\right\rbrace,\label{dpp}\\
{\delta c^2\over c^2}&=&{\gamma-1\over 1-\M^2}\left({\delta f\over c^2}-\M^2{ \delta h}
-\M^2{\delta S}\right),\label{dcc}\\
{\delta p\over\gamma p}&=&{iv\over\omega c^2}\left({\p \delta f\over\p z}-{i\omega\over v}\delta f\right).
\end{eqnarray}
The expression of the perturbations of transverse velocity and vorticity are simply proportional to $\delta f$ and $\delta S$, as deduced from Eqs. (A15-A18) of YF08:
\begin{eqnarray}
\delta v_x &=& {k_x\over\omega}\delta f,\\
\delta w_y&=&{ik_x\over v}\;\;{c^2\over \gamma}\delta S.\label{vorticityy}
\end{eqnarray}
In a uniform flow, the perturbations of energy density $\delta f^S$ and mass flux $\delta h^S$ associated to advected perturbations of entropy/vorticity produced by the shock are given by Eqs.~(C5-C6) in FGSJ07:
\begin{eqnarray}
{\delta f^S\over c^2}&=&{1-\M^2\over 1-\mu^2\M^2}\;\;{\delta S\over \gamma}\label{deffS},\\
\delta h^S&=&{\mu^2\over c^2}\delta f^S-\delta S.\label{defhS}
\end{eqnarray}
The energy density $\delta f^\pm$ and mass flux $\delta h^\pm$ in the acoustic waves are deduced from Eqs.~(C7-C8) of FGSJ07:
\begin{eqnarray}
\delta f^{\pm}&=& {1\over2}f \pm {{\cal M}c^2\over2\mu}(\delta h+\delta S)-
{1\pm\mu\M\over2}\delta f^S
 ,\label{deffpm}\\
\delta h^{\pm}&=&\pm {\mu\over\M}{\delta f^{\pm}\over c^2}.\label{defhpm}
\end{eqnarray}
The perturbations of pressure $\delta p^\pm$ and energy density $\delta f^\pm$ of an acoustic wave in a uniform flow are related by:
\begin{eqnarray}
{\delta p^\pm\over\gamma p}&=&{\delta f^\pm\over c^2}\;\;{1\mp\mu\M\over1-\M^2}.
\end{eqnarray}

\section{WKB approximation}

Owing to the adiabatic hypothesis, the vertical structures of both entropy and vorticity perturbations can be integrated and the differential system can be reduced to a second order differential equation, similar to Foglizzo (2001) when considering the perturbation of energy density $\delta f$ and the variable $X$ defined by:
\begin{eqnarray}
{\dd X\over \dd z}&\equiv& {v\over 1-\M^2},\\
\delta h&=&{1\over i\omega}\;\;{\p \delta f\over\p X}+ {\delta f\over c^2}-\left({1\over\M^2}+\gamma-1\right)
{\delta S\over\gamma},\label{hX}\\
\left\lbrace {\p^2\over\p X^2}+\left({\omega\mu\over v c}\right)^2\right\rbrace\left(
{\delta f\over i\omega}\e^{\int_{\rm sh}{i\omega\over c^2}\dd X}\right)&=&
{\delta S_{\rm sh}\over\gamma}{\p\over\p X}\left(
{1-\M^2\over\M^2}\e^{\int_{\rm sh}{i\omega\over v^2}\dd X}\right).\label{difwkb}
\end{eqnarray}
This equation is formally identical to Eq.~(B32) of Foglizzo (2001) (with $\delta K=0$). This formulation is particularly useful to define a WKB approximation $\delta f_{\rm wkb}^\pm$ of the acoustic solutions of the homogeneous equation in the simplest manner, satisfying a WKB criterion deduced from the left hand side of Eq.~(\ref{difwkb}):
\begin{eqnarray}
{\omega\mu\over c}&\gg&(1-\M^2){\p\log\over\p z}\left({\mu^2\over v^2c^2}\right),\\
\delta f_{\rm wkb}^\pm&\equiv & \delta f_{\rm sh}^\pm{c\over c_{\rm sh}}
\left({\mu_{\rm sh}\M\over \mu\M_{\rm sh}}\right)^{1\over2}
\e^{\int_{\rm sh}{i\omega\over c}{\M\mp\mu\over1-\M^2}\dd z},\label{fwkb}\\
\delta h_\pm&\sim&\pm{\mu\over\M}\;\; {\delta f_\pm\over c^2},\label{hf}
\end{eqnarray}
where $\delta f_{\rm sh}^\pm$ is an arbitrary normalization constant.
The acoustic solution satisfying the lower boundary condition defines the reflection coefficient 
${\cal R}_\nabla$, such that $\delta f_0=\delta f_++{\cal R}_\nabla \delta f_-$ at $z=z_{\rm sh}$.

\section{Compact approximation}

\subsection{Dispersion relation}

The dispersion relation is obtained as in Foglizzo \& Tagger (2000), Foglizzo (2002).
In the specific case of our toy model in the compact approximation, some coupling constants ${\cal Q}'_\nabla$ , 
${\cal R}'_\nabla$ can be defined immediately above the coupling layer ($z\sim z_\nabla$) by
\begin{eqnarray}
\delta f_\nabla^-={\cal Q}'_\nabla \delta f_\nabla^S+{\cal R}'_\nabla \delta f_\nabla^+.
\end{eqnarray}
The value of perturbations at $z=z_{\rm sh}$ and $z=z_\nabla$ are related through their vertical wavenumber $k_z^\pm$ defined by Eq.~(\ref{ky}) and $k_z^0\equiv\omega/v$.
\begin{eqnarray}
\delta f_\sh&=&\delta f_\sh^S+\delta f_\sh^++\delta f_\sh^-,\\
\delta f_\sh^S&=&\e^{ik_z^0\Deltaz}\delta f_\nabla^S,\\
\delta f_\sh^+&=&\e^{ik_z^+\Deltaz}\delta f_\nabla^+,\\
\delta f_\sh^-&=&\e^{ik_z^-\Deltaz}\delta f_\nabla^-.
\end{eqnarray}
The coupling constants ${\cal Q}_\nabla$ and ${\cal R}_\nabla$ can thus be decomposed into a contribution of advection/propagation, and a contribution of the coupling through the compact layer:
\begin{eqnarray}
{\cal Q}_\nabla&=&{\cal Q}'_\nabla\e^{i\omega\tau_{\cal Q}},\\
{\cal R}_\nabla&=&{\cal R}'_\nabla\e^{i\omega\tau_{\cal R}},
\end{eqnarray}
with
\begin{eqnarray}
\tau_{\cal Q}&\equiv& (k_z^--k_z^0){\Deltaz\over\omega}=
\tau_{\rm aac}{1+\mu_{\rm in}\M_{\rm in}\over1+\M_{\rm in}},\\
\tau_{\cal R}&\equiv&(k_z^--k_z^+){\Deltaz\over\omega}
={\Deltaz\over c_{\rm in}}\;\;{2\mu_{\rm in}\over1-\M_{\rm in}^2}.
\end{eqnarray}
Note that the quantities $\tau_{\cal Q}$ and $\tau_{\cal R}$ may be directly interpreted as timescales only when the frequency $\omega$ is real and high enough to correspond to propagating acoustic waves ($\mu_{\rm in}^2>0$).
The global efficiencies ${\cal Q}_0$ and ${\cal R}_0$ are defined by ${\cal Q}_0\equiv {\cal Q}_{\rm sh}{\cal Q}'_\nabla$ and ${\cal R}_0\equiv {\cal R}_{\rm sh}{\cal R}'_\nabla$.

\subsection{Reflection of acoustic waves: calculation of ${\cal R}_\nabla$}

Let an acoustic wave propagate in the direction of the flow towards the region of gradient. It produces
an acoustic reflection $\delta f_{\rm in}^-$ and a transmitted acoustic wave $\delta f_{\rm out}^+$
\begin{eqnarray}
\delta f_{\rm in}^++\delta f_{\rm in}^-&=&\delta f_{\rm out}^+,\label{consf}\\
\delta h_{\rm in}^++\delta h_{\rm in}^-&=&\delta h_{\rm out}^+.\label{consg}
\end{eqnarray}
The system (\ref{consf}-\ref{consg}) is thus equivalent to
\begin{eqnarray}
\delta f_{\rm in}^++\delta f_{\rm in}^-&=&\delta f_{\rm out}^+,\\
\delta f_{\rm in}^+-\delta f_{\rm in}^-&=&{\mu_{\rm out}\M_{\rm in}c_{\rm in}^2
\over \mu_{\rm in}\M_{\rm out}c_{\rm out}^2}\delta f_{\rm out}^+.
\end{eqnarray}
The reflexion coefficient ${\cal R}_\nabla$ in Eq.~(\ref{calR}) is deduced.

\subsection{Advective-acoustic coupling: calculation of ${\cal Q}_\nabla$}

At low frequency, the acoustic perturbations produced by the advection of the entropy/vorticity perturbation through the gradient satisfy the conservation of mass and energy:
\begin{eqnarray}
\delta f_{\rm in}^S+\delta f_{\rm in}^-&=&\delta f_{\rm out}^S+\delta f_{\rm out}^+,\label{consfS}\\
\delta h_{\rm in}^S+\delta h_{\rm in}^-&=&\delta h_{\rm out}^S+\delta h_{\rm out}^+.\label{consgS}
\end{eqnarray}
Using Eq.~(\ref{defhpm}), 
\begin{eqnarray}
\delta f_{\rm in}^--\delta f_{\rm out}^+&=&\delta f_{\rm out}^S-\delta f_{\rm in}^S,\label{confS2}\\
{\mu_{\rm in}\over\M_{\rm in}}\;\;{\delta f_{\rm in}^-\over c_{\rm in}^2}
+{\mu_{\rm out}\over\M_{\rm out}}\;\;{\delta f_{\rm out}^+\over c_{\rm out}^2}&=&
\mu_{\rm in}^2{\delta f_{\rm in}^S\over c_{\rm in}^2}-
\mu_{\rm out}^2{\delta f_{\rm out}^S\over c_{\rm out}^2}
.\label{consgS2}
\end{eqnarray}
thus, the advective-acoustic coupling ${\cal Q}_\nabla$ is described by Eq.~(\ref{calQS}).
$\delta f_{\rm out}^+$ is deduced from $\delta f_{\rm in}^-$ by exchanging the subscripts 'in' and 'out', $\mu_{\rm in}/\M_{\rm in}$ and $-\mu_{\rm out}/\M_{\rm out}$.

\section{Local acoustic emissivity}

Expressing the local pressure emissivity $a_S$ characterizing to the advectic-acoustic coupling due to the gradients of the flow is most easily obtained by writing the differential equation satisfied by pressure perturbations $\delta p$. A pressure perturbation in an adiabatic flow satisfies the acoustic wave equation with a source term associated to the inhomogeneity of the flow. The differential equation established by Foglizzo (2001) in a spherical accelerated flow is formally the same in a decelerated flow. The differential system can be written as follows:
\begin{eqnarray}
\left\lbrace{\p^2\over\p z^2}+a_1{\p\over\p z}+a_2\right\rbrace
\left({1-\M^2\over\M^2}\;\;{\delta p\over p}\right)
=
a_S\delta S,
\end{eqnarray}
with
\begin{eqnarray}
\delta S&=&\delta S_{\rm sh}\e^{\int_{\rm sh}{i\omega\over v}\dd z},\\
\Delta &\equiv&k_x^2v^2+\omega^2-v^3{\p\over\p z}\;{i\omega\over v^2},\\
a_1&=&-{i\omega\over c}\;\;{2\M\over 1-\M^2}-{\p\log\over\p z}\;{\Delta\over v^3},\\
a_2&=&{1\over 1-\M^2}\left(
{\omega^2\over c^2}-k_x^2+v{\p\over\p z}\;{i\omega^2\over v^2}\right)
+{\Delta\over v^3}\;{\p\over\p z}\left({i\omega v^2\over\Delta}\;\;
{1+\M^2\over 1-\M^2}\right),\\
a_S&=&-{\Delta\over v^3}{\p\over\p z}\left\lbrack{c^2\over\Delta}\left(
i\omega -2v{\p\log\M\over\p z}\right)
\right\rbrack.
\end{eqnarray}
Let $\delta p_\pm$ be the acoustic solutions of the homogeneous equation which propagate outward (-) or inward (+), and let $\delta p_0\equiv\delta p_++{\cal R}_\nabla\delta p_-$ be the acoustic solution which satisfies the lower boundary condition at $r_{\rm bc}$. The general solution satisfying the lower boundary condition is written as
\begin{eqnarray}
{\delta p\over\delta S_{\rm sh}} =\delta p_-\int_{\rm bc} {a_S\over W_p}
\e^{\int_{\rm sh}{i\omega\over v}\dd z}
{1-\M^2\over\M^2}{\delta p_0\over p}\dd z
-\delta p_0\left(\beta+\int_{\rm bc} {a_S\over W_p}
\e^{\int_{\rm sh}{i\omega\over v}\dd z}{1-\M^2\over\M^2}
\;\;{\delta p_-\over p}\dd z\right),
\end{eqnarray}
where $W$ contains the normalization of the amplitudes of $\delta p_\pm$:
\begin{eqnarray}
W_p&\equiv&{1-\M^2\over\M^2}\left\lbrack
{\delta p_0\over p}\;{\p\over\p z}\left({1-\M^2\over\M^2}\;\;{\delta p_-\over p}\right)
-{\delta p_-\over p}\;{\p\over\p z}\left({1-\M^2\over\M^2}\;\;{\delta p_0\over p}\right)\right\rbrack,\\
{\p\log W_p\over\p z}&=&-a_1.
\end{eqnarray}
The WKB approximation of the pressure perturbation deduced from Eq.~(\ref{fwkb}) is
\begin{eqnarray}
{\delta p^\pm_{\rm wkb}\over\gamma p}&\sim& 
{1\mp\mu\M\over1-\M^2}\;\;{f_{\rm wkb}^\pm\over c^2},\\
&\sim& {f_{\rm sh}^\pm\over c_{\rm sh}^2}{c_{\rm sh}\over c}
\left({\M\mu_{\rm sh}\over\mu\M_{\rm sh}}\right)^{1\over2}
{1\mp\mu\M\over1-\M^2}\e^{\int_{\rm sh}{i\omega\over c}{\M\mp\mu\over1-\M^2}\dd z},\\
W_p&\sim&-2i\gamma^2{\mu_{\rm sh}\over\M_{\rm sh}}\;\;{f_{\rm sh}^+f_{\rm sh}^-\over \omega c_{\rm sh}^2}\;\;
{\Delta\over v^3}
\e^{\int_{\rm sh}{i\omega\over c}{2\M\over1-\M^2}\dd z}.
\end{eqnarray}
Thus
\begin{eqnarray}
{a_S\over W_p}&=&-{i\omega\M_{\rm sh}c_{\rm sh}^2\over2f_{\rm sh}^+f_{\rm sh}^-\gamma^2 \mu_{\rm sh}}
\e^{\int_{\rm sh} {i\omega\over v}{2\M^2\over 1-\M^2}\dd z}
{\p\over\p z}\left\lbrack{c^2\over\Delta}\left(
i\omega -2v{\p\log\M\over\p z}\right)
\right\rbrack.
\end{eqnarray}
The derivative of the pressure perturbation is:
\begin{eqnarray}
{\p\over\p z}\;{\delta p\over\delta S_{\rm sh}} = {\p\over\p z}\delta p_-\left(\int_{\rm bc} {a_S\over W_p}
\e^{\int_{\rm sh}{i\omega\over v}\dd z}
{1-\M^2\over\M^2}\;\;{\delta p_0\over p}\dd z\right)
-{\p\over\p z}\delta p_0\left(\beta+\int_{\rm bc} {a_S\over W_p}
\e^{\int_{\rm sh}{i\omega\over v}\dd z}{1-\M^2\over\M^2}\;\;
{\delta p_-\over p}\dd z\right).
\end{eqnarray}
The outgoing pressure perturbation is deduced from the WKB approximation:
\begin{eqnarray}
\delta p&=&\delta {\tilde p}_-+\delta {\tilde p}_+,\\
{\p\over\p z}\delta p&=&{i\omega\over c}\left(
{\M+\mu\over1-\M^2}\delta {\tilde p}_-+{\M-\mu\over1-\M^2}\delta {\tilde p}_+\right).
\end{eqnarray}
Each component $\delta \tilde p^\pm$ is deduced from $\delta p$ and its derivative according to:
\begin{eqnarray}
\delta \tilde p_\pm={1\over2}\left\lbrace
\left(1\pm{\M\over\mu}\right)\delta p\mp {c\over i\omega\mu}(1-\M^2){\p\over\p z}\delta p\right\rbrace.
\end{eqnarray}
The value of $\beta$ is deduced from the outer boundary condition $\delta\tilde p_-=0$, and the general solution satisfying the boundary conditions is thus simply:
\begin{eqnarray}
{\delta p\over\delta S_{\rm sh}} = \delta p_-\int_{\rm bc} {a_S\over W_p}
\e^{\int_{\rm sh}{i\omega\over v}\dd z}
{1-\M^2\over\M^2}\;\;{\delta p_0\over p}\dd z
-\delta p_0\int_{\rm sh} {a_S\over W_p}
\e^{\int_{\rm sh}{i\omega\over v}\dd z}{1-\M^2\over\M^2}
\;\;{\delta p_-\over p}\dd z.
\end{eqnarray}
The acoustic feedback produced by the advection of the entropy-vorticity perturbation is 
\begin{eqnarray}
{\cal Q}_\nabla
&=&(1-\mu_{\rm sh}\M_{\rm sh}){\delta p_{\rm sh}\over p\delta S_{\rm sh}},\\
&=&
\left(1+{k_x^2v_{\rm sh}^2\over \omega^{2}}\right)
\left(
1-{\cal R}_\nabla-{1+{\cal R}_\nabla\over\mu_{\rm sh}\M_{\rm sh}}
\right)\nonumber\\
&&\times\int_{\rm bc}^{\rm sh}
{{1-\M^2\over\M^2}\;\;{\delta p_0\over p}\over  \left({1-\M^2\over\M^2}\;\;{\delta p_0\over p}\right)_{\rm sh}}
\e^{\int_{\rm sh}
{i\omega\over v}{1+\M^2\over1-\M^2}\dd z}
{i\omega\over 2c_{\rm sh}^2}\;{\p\over\p z}\left\lbrack{
i\omega-v{\p\log\M^2\over\p z}\over
k_x^2\M^2+{\omega^2\over c^2}-v\M^2{\p\over\p z}\;{i\omega\over v^2} }
\right\rbrack\dd z\label{qnabla},
\end{eqnarray}
from which we deduce Eqs.~(\ref{qnabgrad}-\ref{gradients}).


\begin{thebibliography}{}

\bibitem[BW85]{BW85}
Bethe, H.A. \& Wilson, J.R. 1985, \apj, 295, 14

\bibitem[2003]{bmd03}
Blondin, J.M, Mezzacappa, A., \& DeMarino, C. 2003,  \apj, 584, 971

\bibitem[2003]{bm06}
Blondin, J.M, \& Mezzacappa, A. 2006,  \apj, 642, 401

\bibitem[Blondin \& Mezzacappa (2007)]{BM07}
Blondin, J.M, \& Mezzacappa, A. 2007, Nature, 445, 58

\bibitem[Blondin \& Shaw (2007)]{BS07}
Blondin, J.M, \& Shaw, A. 2007, \apj, 656, 366

\bibitem[BRJK03]{BRJK03}
Buras, R., Rampp, M., Janka, H.-T., \& Kifonidis, K. 2003, \prl, 90, 241101

\bibitem[BHF95]{BHF95}
Burrows, A., Hayes, J., \& Fryxell, B.A. 1995, \apj, 450, 830

\bibitem[Burrows et al. (2006)]{Burrows+06}
Burrows, A., Livne, E., Dessart, L., Ott, C. D., 
\& Murphy, J. 2006, ApJ, 640, 878

\bibitem[Burrows et al. (2007)]{Burrows+07}
Burrows, A., Livne, E., Dessart, L., Ott, C. D., 
\& Murphy, J. 2007, ApJ, 655, 416

\bibitem[2001]{f01}
Foglizzo, T. 2001, \aap, 368, 311

\bibitem[2002]{f02}
Foglizzo, T. 2002, \aap, 392, 353

\bibitem[2000]{ft00}
Foglizzo, T., \& Tagger, M. 2000, \aap, 363, 174

\bibitem[2006]{fgr06}
Foglizzo, T., Galletti, P., \& Ruffert, M. 2005, \aap, 435, 397

\bibitem[2006]{fsj06}
Foglizzo, T., Scheck, L., \& Janka, H.T. 2006, \apj, 652, 1436 (FSJ06)

\bibitem[Foglizzo et al.(2007)]{Foglizzo+07}
Foglizzo, T., Galletti, P., Scheck, L., \& Janka, H.-Th. 2007, ApJ, 654, 1006 (FGSJ07)

\bibitem[Iwakami et al.(2008)]{Iwakami+08} 
Iwakami, W., Kotake, K., Ohnishi, N., Yamada, S., \& Sawada, K. 2008, 
ApJ, 678, 1207 

\bibitem[JM96]{JM96}
Janka, H.-T. \& M\"uller, E. 1996, \aap, 306, 167

\bibitem[Laming (2007)]{Laming07}
Laming, J. M. 2007, ApJ, 659, 1449

\bibitem[Laming (2008)]{Laming08}
Laming, J. M. 2008, ApJ, Erratum in press

\bibitem[L01]{L01}
Liebend\"orfer M. et al. 2001, \prd, 63, 103004

\bibitem[MC]{mc77}
Marble, F.E., \& Candel, S.M., 1977, Journal of Sound and Vibration, 55, 225

\bibitem[Marek \& Janka (2007)]{MJ07}
Marek, A., \& Janka, H.-Th. 2007, submitted to \apj (arXiv: 0708.3372)

\bibitem[MB]{MB08}
Murphy, J.W., \& Burrows, A. 2008, \apj, 688, 1159

\bibitem[OKY]{oky06}
Ohnishi, N., Kotake, K., \& Yamada, S. 2006, \apj, 641, 1018

\bibitem[Foglizzo (2008)]{F08}
Sato, J., Foglizzo, T., \& Fromang 2008, \apj, submitted (paper~II)

\bibitem[Scheck et al.(2004)]{Scheck+04}
Scheck, L., Plewa, T., Janka, H.-Th., Kifonidis, K., \& M\"{u}ller, E. 
2004, Phys. Rev. Lett., 92, 011103 

\bibitem[SK]{skjm06}
Scheck, L., Kifonidis, K., Janka, H.T., \& M\"uller, E. 2006, \aap, 457, 963

\bibitem[Scheck et al.(2008)]{Scheck+08}
Scheck, L., Janka, H.-Th., Foglizzo, T., \& Kifonidis, K. 
2008, \aap, 477, 931 

\bibitem[Weinberg \& Quataert (2008)]{WQ08}
Weinberg, N.N., Quataert, E. 2008, MNRAS 387, L64

\bibitem[Yamasaki \& Foglizzo (2008)]{YF08} 
Yamasaki, T. \& Foglizzo, T. 2008, ApJ, 679, 607

\end{thebibliography}
\end{document}